\documentclass[aps,pre,onecolumn,notitlepage,superscriptaddress]{revtex4-2}

\usepackage[margin=0.7in, a4paper]{geometry}                	 
\usepackage[colorlinks]{hyperref}
\usepackage[utf8]{inputenc}

\usepackage{calc}

\usepackage[explicit,calcwidth]{titlesec}
\usepackage{braket}
\usepackage{amsmath}
\usepackage{amsthm}
\usepackage{amssymb}
\usepackage{amsfonts}
\usepackage{graphicx}				 
\usepackage[makeroom]{cancel}

\newcommand{\rmi}{\mathrm{in}}
\newcommand{\rmf}{\mathrm{fin}}
\newcommand{\tr}{\mathrm{Tr}}

\begin{document}

\title{Stochastic entropy production: Fluctuation relation and irreversibility mitigation in non-unital quantum dynamics}

\author{Eliana Fiorelli}
\affiliation{Instituto de F\'isica Interdisciplinar y Sistemas Complejos (IFISC), UIB–CSIC \\
UIB Campus, Palma de Mallorca, E-07122, Spain}

\author{Stefano Gherardini}
\affiliation{Istituto Nazionale di Ottica – CNR, Area Science Park, Basovizza, I-34149 Trieste, Italy}
\affiliation{LENS, University of Florence, via Carrara 1, I-50019 Sesto Fiorentino, Italy}
\affiliation{The Abdus Salam International Center for Theoretical Physics (ICTP), Strada Costiera 11, I-34151 Trieste, Italy}

\author{Stefano Marcantoni}
\affiliation{School of Physics and Astronomy, University of Nottingham, Nottingham, NG7 2RD, UK}
\affiliation{Centre for the Mathematics and Theoretical Physics of Quantum Non-Equilibrium Systems, University of Nottingham, Nottingham, NG7 2RD, UK}
\affiliation{Mathematics Area, SISSA, Via Bonomea 265, I-34136 Trieste, Italy}

\begin{abstract}
In this work, we study the stochastic entropy production in open quantum systems whose time evolution is described by a class of non-unital quantum maps. In particular, as in {\it Phys.~Rev.~E {\bf 92}, 032129 (2015)}, we consider Kraus operators that can be related to a nonequilibrium potential. This class accounts for both thermalization and equilibration to a non-thermal state. Unlike unital quantum maps, non-unitality is responsible for an unbalance of the forward and backward dynamics of the open quantum system under scrutiny. Here, concentrating on observables that commute with the invariant state of the evolution, we show how the non-equilibrium potential enters the statistics of the stochastic entropy production. In particular, we prove a fluctuation relation for the latter and we find a convenient way of expressing its average solely in terms of relative entropies. Then, the theoretical results are applied to the thermalization of a qubit with non-Markovian transient, and the phenomenon of irreversibility mitigation, introduced in {\it Phys.~Rev.~Research {\bf 2}, 033250 (2020)}, is analyzed in this context. 
\end{abstract}

\maketitle

\section{Introduction}

The entropy production is a key quantity in the study of open quantum systems (quantum systems in interaction with the external environment). On the one hand, it identifies the amount of energy that is irreversibly lost in terms of heat during the interaction process, and, on the other hand, it quantifies the degree of irreversibility of the resulting dynamics~\cite{BenattiBook2003,EspositoNJP2010,EspositoPRB2015,Marcantoni2017entropy,BatalhaoBook2019,LandiRMP2021}. In this paper, we adopt the point of view of stochastic quantum thermodynamics, so that the entropy production is defined as a stochastic variable, whose probability distribution and statistical moments can be computed by resorting to the Tasaki-Crooks formalism~\cite{Crooks99,Tasaki00,TalknerJPA07,CrooksPRA2008}. According to such a formalism, the stochastic entropy production quantifies how much the joint probability distribution of two measurement outcomes at different times along a quantum evolution differs from the corresponding distribution in the time-reversed dynamics.

Since we are dealing with open quantum systems, the dynamics is not unitary and the definition of a time-reversed path is not straightforward. If the system and its environment are uncorrelated at time $t=0$, the evolution at later times is described in general by a one-parameter family of Completely Positive and Trace Preserving (CPTP) maps~\cite{BreuerBook2007}. Many results in quantum thermodynamics deal with the so-called \emph{unital} maps, namely with the special class of time evolutions that preserve the maximally-mixed state (multiple of the identity). In fact, for unital maps there is a natural notion of inverse dynamics given by the dual quantum map. Moreover, in this case the entropy production is known to depend only on the initial and final quantum states of the nonequilibrium process under scrutiny, not on the full history. As a result, on average the entropy production is provided by the difference of the von Neumann entropies in the final and initial states.
In this regard, fluctuation relations and the statistics of thermodynamic quantities in unital quantum maps have been studied for instance in~\cite{Rastegin13,Albash13,Rastegin14}. It is worth mentioning in particular the case of unitary dynamics interspersed by projective measurements that also belong to this class~\cite{Campisi10,Campisi11b,Gherardini18,Giachetti20,GherardiniPRE2021,GherardiniReview2022}.

However, a lot of physically interesting phenomena are described by non-unital maps, like thermalization, dissipation to a non-thermal state, and even feedback-controlled mechanisms as e.g.~those of Maxwell demons.
In this setting, the definition of a time-reversed protocol is less straightforward; however, a few proposals appeared in the literature~\cite{CrooksPRA2008,Aurell15,ChiribellaPRR2021,Goold2021fluctuation} so that the Tasaki-Crooks formalism equally applies to the thermodynamics of non-unital quantum maps. It turns out that, in general, for non-unital dynamics the entropy production has a more complicated structure than for the unital ones, both on average and at the stochastic level, where fluctuations are involved. Therefore, it is important to investigate whether simpler and physically informative results can still be proved for particular classes of non-unital maps (maybe with some restrictions on the initial state and/or on the measured observables). On top of this, since the formalism of quantum dynamical maps allows to consider both Markovian and non-Markovian dynamics (in the sense of~\cite{Chruscinski2014degree}), one could also ask whether the degree of non-Markovianity plays a relevant role.
In fact, despite the interest in non-Markovian effects related to quantum thermodynamics has increased in the last decade, and a few results have been established for average thermodynamic quantities, much less is known at the level of fluctuations.
A non-exhaustive list of references about quantum thermodynamics in these contexts is~\cite{Sagawa08,Sagawa10,Morikuni11,Funo13,Rastegin14,Lorenzo15,Niedenzu16,Campisi17,Marcantoni2017entropy,Masuyama18,Ramezani18,Pal19,Strasberg19,Colla22,Gherardini2020irreversibility,HernandezGomez20,GherardiniPRE2021,HernandezGomezPRXQuantum22}.

Motivated by the previous discussion, in this paper we study the stochastic entropy production in a class of non-unital quantum dynamics that describes, for instance, thermalization with a non-Markovian transient. In Section~\ref{sec:stochastic_entropy}, we describe the Tasaki-Crooks formalism by giving the definition of stochastic entropy production. In Section~\ref{sec:fluctuation_relation}, we present two main results. In particular, we show that, for the considered class of dynamics and for the set of observables commuting with the invariant state of the map, i) the stochastic entropy production obeys a fluctuation relation and ii) the average entropy production can always be written solely in terms of quantum relative entropies. We then comment on the difference between our results and the related work done in Ref.~\cite{Manzano15,ManzanoPRX2018}. In Section~\ref{sec:qubit_thermalization}, we briefly introduce the concept of quantum non-Markovianity and then we discuss a thermalizing qubit dynamics with non-Markovian transient, computing explicitly the stochastic entropy production and its first two moments. We also compute the rate of variation of the average and of the variance of the stochastic entropy production. This allows us to find our third main result. Indeed, in Section~\ref{sec:irreversibility_mitigation}, as in the unital setting, iii) we find a parameter range such that, due to non-Markovianity, the average entropy production and its variance are simultaneously decreasing in the transient. This is the phenomenon we called \emph{irreversibility mitigation}~\cite{Gherardini2020irreversibility}. Finally, in Section~\ref{sec:conclusions}, we present conclusions and perspectives for future investigations.

\section{Stochastic entropy production}
\label{sec:stochastic_entropy}

Let us consider a $d$-level open quantum system whose evolution up to time $t$ is described by a CPTP map $\Lambda_t$. We consider the latter having a unique positive-definite invariant state $\pi=\sum_{i}\pi_i |\pi_i\rangle\!\langle\pi_i|>0$, with probabilities $\pi_i\in[0,1]$ and $\sum_{i}\pi_{i}=1$, such that the map is in general \emph{non-unital}. We require the uniqueness of $\pi$ because, in the following, we use a notion of time-reversed dynamics that is based on the invariant state.

In order to define the stochastic entropy production, which is the quantity of interest in this paper, we need to consider two distinct procedures, namely a forward and backward protocol, in accordance with the Tasaki-Crooks formalism~\cite{Crooks99,Tasaki00,TalknerJPA07,CrooksPRA2008}. In particular, we resort to the two-point measurement scheme (TPM)~\cite{Kafri12,Cimini20,AguilarPRA2022}; this choice stems also from our intention to compare the results we are going to present with the ones determined in~\cite{Manzano15,Gherardini2020irreversibility}. In the time interval $[0,\tau]$, the TPM consists in applying two projective measurements, at the initial and at the final time instants, of the two observables $\mathcal{O}_{\mathrm{in}}$ and $\mathcal{O}_{\mathrm{fin}}$ (Hermitian operators)~\cite{GherardiniQST2018,ManzanoPRX2018}. The spectral decomposition of the observables is $\mathcal{O}_{\mathrm{in}} = \sum_m a_m^{\mathrm{in}}\Pi_m^{\mathrm{in}}$ and $\mathcal{O}_{\mathrm{fin}}=\sum_k a_k^{\mathrm{fin}}\Pi_k^{\mathrm{fin}}$, where each $\lbrace\Pi\rbrace$ identifies the set of projectors associated to each set of eigenvalues/measurement outcomes $\lbrace a \rbrace$. The forward process is defined by the following sequence of operations: 
\begin{equation}\label{eq:forward_process}
\begin{Large}
\rho_0 \underbrace{\longrightarrow}_{\Pi_m^{\mathrm{in}}} \rho_{\mathrm{in}} \underbrace{\longrightarrow}_{\Lambda_\tau} \rho_{\tau} \underbrace{\longrightarrow}_{\Pi_k^{\mathrm{fin}}} \rho_{\mathrm{fin}}
\end{Large}
\end{equation}
where, after the first projective measurement, the state of the system is on average $\rho_{\mathrm{in}}=\sum_{m}p(a_{m}^{\mathrm{in}})\Pi_{m}^{\mathrm{in}}$ with $p(a_m^{\mathrm{in}})= \mathrm{Tr}[\Pi_{m}^{\mathrm{in}}\rho_0 \Pi_{m}^{\mathrm{in}}]=\mathrm{Tr}[\rho_0\Pi_{m}^{\rm in}]$. Applying the second projective measurement to the state $\rho_{\tau}\equiv\Lambda_{\tau}[\rho_{\mathrm{in}}]$ after the open quantum dynamics, one gets $\rho_{\tau}=\sum_{k}p(a_{k}^{\mathrm{fin}})\Pi_{k}^{\mathrm{fin}}$ with $p(a_{k}^{\mathrm{fin}})= \mathrm{Tr}[\Pi_{k}^{\mathrm{fin}}\Lambda_{\tau} (\rho_\mathrm{in}) \Pi_{k}^{\mathrm{fin}} ] = \mathrm{Tr}[\Lambda_{\tau}(\rho_{\rm in})\Pi_{k}^{\mathrm{fin}}]$. Hence, the joint probability that the measurement outcome $a_{m}^{\mathrm{in}}$ is followed by the measurement outcome $a_{k}^{\mathrm{fin}}$ in the forward process is
\begin{equation}\label{eq:forward_joint}
p_{\mathrm{F}}(a_k^{\mathrm{fin}}, a_m^{\mathrm{in}})= \mathrm{Tr}[\Pi_{k}^{\mathrm{fin}}\Lambda_{\tau} ( \Pi_{m}^{\mathrm{in}}\rho_\mathrm{0} \Pi_{m}^{\mathrm{in}}) ] = \mathrm{Tr}[\Pi_{k}^{\mathrm{fin}}\Lambda_{\tau} ( \Pi_{m}^{\mathrm{in}}) ]p(a_{m}^{\mathrm{in}})\, .
\end{equation}
Conversely, the backward process consists in the following set of operations:
\begin{equation}\label{eq:backward_process}
\begin{Large}
\tilde{\rho}_{\mathrm{fin}} \underbrace{\longrightarrow}_{\tilde{\Pi}_k^{\mathrm{fin}}} \tilde{\rho}_{\tau} \underbrace{\longrightarrow}_{\tilde{\Lambda}_{\tau}} \tilde{\rho}_{\mathrm{in}} \underbrace{\longrightarrow}_{\tilde{\Pi}_m^{\mathrm{in}}} \tilde{\rho}_{\mathrm{0}}\,.
\end{Large}
\end{equation}
More explicitly, following~\cite{Manzano15}, the main ingredient of the backward process is the dual map $\tilde{\Lambda}_t$. Given the invariant state $\pi$ and the Kraus representation of $\Lambda_t$, i.e., $\Lambda_t(\cdot)=\sum_{\ell}E_{\ell}(t)(\cdot)E_{\ell}^{\dagger}(t)$ where $\{E_\ell\}_{\ell=1}^{d^2}$ is the set of Kraus operators and $d$ is the dimension of the system's Hilbert space, the dual map is defined as $\tilde{\Lambda}_t(\cdot) = \sum_{\ell}\tilde{E}_{\ell}(t)(\cdot)\tilde{E}_{\ell}^{\dagger}(t)$, where 
$\tilde{E}_{\ell}(t) \equiv \Theta \pi^{\frac{1}{2}}E_{\ell}^{\dagger}(t)\pi^{-\frac{1}{2}}\Theta^{\dagger}$. Here $\Theta$ is the antiunitary time-reversal operator obeying the relations $\Theta^\dag \Theta=\Theta \Theta^\dag = \mathbb{I}$. By construction, $\tilde{\Lambda}_t$ is CPTP and its invariant state is $\tilde{\pi}=\Theta\pi\Theta^{\dagger}$, namely the time-reversed version of $\pi$.
Additionally, in order to fully specify the operations in Eq.~(\ref{eq:backward_process}), we also set the first [second] measurement of the backward process to be the time-reversal of the second [first] measurement of the forward process, i.e., $\tilde{\Pi}_{k}^{\mathrm{fin}} \equiv \Theta \Pi_{k}^{\mathrm{fin}} \Theta^\dag$ and $\tilde{\Pi}_{m}^{\mathrm{in}} \equiv \Theta \Pi_{m}^{\mathrm{in}} \Theta^\dag$. With this assumption, the joint probability that in the backward process the measurement outcome $a_{k}^{\mathrm{fin}}$ is followed by the measurement outcome $a_{m}^{\mathrm{in}}$ is
\begin{equation}\label{eq:backward_joint}
p_{\mathrm{B}}(a_m^{\mathrm{in}}, a_k^{\mathrm{fin}})= \mathrm{Tr}[\tilde{\Pi}_{m}^{\mathrm{in}} \tilde{\Lambda}_t( \tilde{\Pi}_{k}^{\mathrm{fin}}\tilde{\rho}_\mathrm{fin} \tilde{\Pi}_{k}^{\mathrm{fin}}) ] = \mathrm{Tr}[\tilde{\Pi}_{m}^{\mathrm{in}}\tilde{\Lambda}_t( \tilde{\Pi}_{k}^{\mathrm{fin}}) ]p(a_{k}^{\mathrm{fin}}).
\end{equation}

The stochastic quantum entropy production is then defined as~\cite{GherardiniQST2018,ManzanoPRX2018,Gherardini2020irreversibility} 
\begin{equation}\label{e_stoc_QEP}
\Delta\sigma(a_k^{\mathrm{fin}}, a_m^{\mathrm{in}}) \equiv \ln{\left[ \frac{p_{\mathrm{F}}(a_k^{\mathrm{fin}}, a_m^{\mathrm{in}})}{p_{\mathrm{B}}( a_m^{\mathrm{in}},a_k^{\mathrm{fin}})} \right] } = \ln[p(a_m^{\mathrm{in}})]-\ln[p(a_k^{\mathrm{fin}})]+\ln{\left[ \frac{p_{\mathrm{F}}(a_k^{\mathrm{fin}}| a_m^{\mathrm{in}})}{p_{\mathrm{B}}( a_m^{\mathrm{in}}|a_k^{\mathrm{fin}})} \right]}.
\end{equation}
If the quantum map $\Lambda_t$ is unital, then $p_{\mathrm{F}}(a_k^{\mathrm{fin}}| a_m^{\mathrm{in}})=p_{\mathrm{B}}( a_m^{\mathrm{in}}|a_k^{\mathrm{fin}})$, such that $\ln{\left[p_{\mathrm{F}}(a_k^{\mathrm{fin}}| a_m^{\mathrm{in}})/p_{\mathrm{B}}( a_m^{\mathrm{in}}|a_k^{\mathrm{fin}})\right]}=0$ and thus $\Delta\sigma(a_k^{\mathrm{fin}}, a_m^{\mathrm{in}})=\ln[p(a_m^{\mathrm{in}})]-\ln[p(a_k^{\mathrm{fin}})]$. In the non-unital case, instead, the forward and backward conditional probabilities are not equal to each other, and in general their expression depends on the details of the quantum map $\Lambda_t$.

\section{Fluctuation relation}
\label{sec:fluctuation_relation}

Even in the non-unital case, one would like to determine if under some assumptions the ratio of conditional probabilities $p_{\mathrm{F}}(a_k^{\mathrm{fin}}|a_m^{\mathrm{in}})/p_{\mathrm{B}}(a_m^{\mathrm{in}}|a_k^{\mathrm{fin}})$ can be expressed in terms of quantities that do \emph{not} depend on time.
As we will show below, this is indeed possible in case the proposal of Ref.~\cite{CrooksPRA2008} is used to define the time-reversal protocol, and such a key aspect allows to derive a fluctuation relation for the stochastic entropy production in non-unital dynamics, like the ones describing thermalization. It is worth noting that other choices of the backward dynamics, as in Refs.~\cite{ChiribellaPRR2021,Goold2021fluctuation}, do not necessarily guarantee that the ratio of conditional probabilities $p_{\mathrm{F}}(a_k^{\mathrm{fin}}|a_m^{\mathrm{in}})/p_{\mathrm{B}}(a_m^{\mathrm{in}}|a_k^{\mathrm{fin}})$, although defined starting from the quantum map $\Lambda_t$, does not depend on time.

To get the fluctuation relation, we make use of the concept of \emph{nonequilibrium potential} -- originally introduced in \cite{Manzano15} -- that is defined in terms of the invariant state $\pi$ of the quantum map $\Lambda_t$. 
Given the spectral decomposition $\pi= \sum_i \pi_i |\pi_i\rangle\!\langle\pi_i|$,
we assign to each projector $|\pi_i\rangle\!\langle\pi_i|$ the nonequilibrium potential term 
\begin{equation}
\Phi_\pi(i) \equiv -\ln[\pi_i]\,.
\end{equation}
Then, we focus on a class of quantum maps in Kraus representation where each Kraus operator $E_\ell$ is a linear combination of ``jump'' operators $|\pi_j\rangle\!\langle\pi_i|$ associated with the \emph{same} change $\Delta \Phi_{\pi}(\ell)$ of the nonequilibrium potential. 
As we are going to show below, this assumption is the \emph{first} of a pair of conditions that are required to express $\ln{\left[p_{\mathrm{F}}(a_k^{\mathrm{fin}}| a_m^{\mathrm{in}})/p_{\mathrm{B}}( a_m^{\mathrm{in}}|a_k^{\mathrm{fin}})\right]}$ as a function of the $\Phi_{\pi}(i)$'s, provided the TPM scheme is applied. 
Following the previous discussion, the single Kraus operator $E_{\ell}(t)$ can be written as
\begin{equation}
E_{\ell}(t) = \sum_{ij}m_{ji}^{(\ell)}(t)
|\pi_j\rangle\!\langle\pi_i|\,,
\end{equation}
with the constraint that the coefficients
\begin{equation}\label{eq:impl_def_Delta_Phi}
m_{ji}^{(\ell)}(t)=0 \,\,\,\forall t \quad \text{if} \quad \Phi_{\pi}(j)-\Phi_{\pi}(i) \neq \Delta \Phi_{\pi}(\ell) \,.
\end{equation}
Hence, all the transitions $(i,j)$ that are contained in the same Kraus operator $E_{\ell}(t)$ are characterized by the same potential $\Delta\Phi_{\pi}(\ell)$; of course, there may be different Kraus operators that are associated to the same value of $\Delta\Phi_{\pi}$. 
With these assumption, the following relations hold true at any time $t$:
\begin{eqnarray}
&& \pi^{-1/2} E_{\ell}(t) = e^{\Delta \Phi_{\pi}(\ell)/2}E_{\ell}(t) \pi^{-1/2} \label{e_comm_Epi1} \\
&& E_{\ell}^\dag(t) \pi^{-1/2}= e^{\Delta \Phi_{\pi}(\ell)/2} \pi^{-1/2} E_{\ell}^\dag(t) \label{e_comm_Epi2}
\end{eqnarray}
since 
\begin{eqnarray}\label{eq:commutatore}
[E_{\ell}(t),\pi^{-1/2}] &=& \sum_{ij}(\pi_i^{-1/2} - \pi_j^{-1/2}) \, m_{ji}^{(\ell)}(t)|\pi_j\rangle\!\langle\pi_i|
= \sum_{ij} \left(1-\left(\frac{\pi_j}{\pi_i}\right)^{-\frac{1}{2}}\right)\pi_i^{-1/2} m_{ji}^{(\ell)}(t)|\pi_j\rangle\!\langle\pi_i| \nonumber \\
&=& \left(1-e^{\Delta \Phi_{\pi}(\ell)/2}\right)E_{\ell}(t)\pi^{-1/2} \,,
\end{eqnarray}
where the last equality in Eq.~(\ref{eq:commutatore}) comes from using Eq.~(\ref{eq:impl_def_Delta_Phi}). One can use the relations~(\ref{e_comm_Epi1}) and~(\ref{e_comm_Epi2}) to compare the forward conditional probability 
\begin{equation}\label{eq:forward}
 p_{\mathrm{F}}(a_k^{\mathrm{fin}}| a_m^{\mathrm{in}}) = \sum_{\ell} \mathrm{Tr}\left[\Pi_k^{\mathrm{fin}} E_{\ell}(t) \Pi_m^{\mathrm{in}} E_{\ell}^{\dagger}(t)\right] 
\end{equation}
with the backward one
\begin{eqnarray}\label{eq:backward}
  p_{\mathrm{B}}(a_m^{\mathrm{in}}|a_k^{\mathrm{fin}}) &=& \sum_{\ell} \mathrm{Tr}\left[\tilde{\Pi}_{m}^{\mathrm{in}} \tilde{E}_{\ell}(t)\tilde{\Pi}_{k}^{\mathrm{fin}}\tilde{E}_{\ell}^{\dagger}(t)\right] = 
  \sum_{\ell} \mathrm{Tr}\left[\Theta \Pi_{m}^{\mathrm{in}} \Theta^\dag \Theta \pi^{1/2} {E}_{\ell}^{\dagger}(t) \pi^{-1/2} \Theta^\dag \Theta {\Pi}_{k}^{\mathrm{fin}} \Theta^\dag \Theta \pi^{-1/2} {E}_{\ell}(t) \pi^{1/2} \Theta^\dag \right] \nonumber \\
&=&  
\sum_{\ell} \mathrm{Tr}\left[\Theta \Pi_m^{\mathrm{in}} e^{\Delta \Phi_{\pi}(\ell)/2} E_{\ell}^{\dagger}(t) \Pi_k^{\mathrm{fin}} e^{\Delta \Phi_{\pi}(\ell)/2} E_{\ell}(t) \Theta^\dag \right]= \sum_{\ell} e^{\Delta \Phi_{\pi}(\ell)}\mathrm{Tr}\left[ E_{\ell}^{\dagger}(t) \Pi_k^{\mathrm{fin}} E_{\ell}(t) \Pi_m^{\mathrm{in}} \right] \nonumber \\
&=&  \sum_{\ell} e^{\Delta \Phi_{\pi}(\ell)}\mathrm{Tr}\left[  \Pi_k^{\mathrm{fin}} E_{\ell}(t) \Pi_m^{\mathrm{in}} E_{\ell}^{\dagger}(t) \right].
\end{eqnarray}
Note that in the second step of the second line we used the property $\mathrm{Tr}\left[\Theta A \Theta^\dag \right] = \mathrm{Tr}\left[ A^\dag\right]$ valid for antiunitary operators (the usual ciclicity of the trace does not hold). This property can be proved from the characterization of antiunitary operators ($\Theta$ antilinear such that $\braket{\Theta \psi, \Theta \varphi}= \overline{\braket{\psi, \varphi}} $) and their adjoint ($\Theta^\dag$ antilinear such that $\braket{\Theta \psi, \varphi}= \overline{\braket{\psi, \Theta^\dag \varphi}}$). 

When taking the ratio $p_{\mathrm{F}}(a_k^{\mathrm{fin}}| a_m^{\mathrm{in}})/p_{\mathrm{B}}( a_m^{\mathrm{in}}|a_k^{\mathrm{fin}})$, one can see from Eqs.~(\ref{eq:forward}) and~(\ref{eq:backward}) that the summation over $\ell$ (different Kraus operators) does not allow simplifications in general. A special case is constituted by unital maps, where all $\Delta \Phi$s are vanishing, so that the ratio of conditional probabilities equals one. Dealing with non-unital maps, one needs a further assumption to express $\ln{\left[p_{\mathrm{F}}(a_k^{\mathrm{fin}}| a_m^{\mathrm{in}})/p_{\mathrm{B}}( a_m^{\mathrm{in}}|a_k^{\mathrm{fin}})\right]}$ only in terms of $\Delta \Phi_{\pi}$. In particular, the \emph{second} assumption we make is that both observables $\mathcal{O}_{\rm in}$ and $\mathcal{O}_{\rm fin}$ commute with the invariant state $\pi$ of the forward process.
In other terms, one needs that $[\mathcal{O}_{\rmi},\pi]=[\mathcal{O}_{\rmf},\pi]=0$, such that the measurements applied at the initial and final times of the quantum process are described by the following projectors
\begin{equation}\label{e_projective_meas}
\Pi_m^{\mathrm{in}} = |\pi_m\rangle\!\langle\pi_m|, \quad \text{and} \quad \Pi_{k}^{\mathrm{fin}} =  |\pi_k\rangle\!\langle\pi_k| \, .
\end{equation}
In this way, a specific transition between $\ket{\pi_m}$ and $\ket{\pi_k}$ is selected by the measurement, and this transition is in turn associated to a single value of the nonequilibrium potential, which can be then extracted from the summation. More explicitly, one can write
\begin{equation}\label{e_SE_noneq}
\ln{\left[ \frac{p_{\mathrm{F}}(a_k^{\mathrm{fin}}| a_m^{\mathrm{in}})}{p_{\mathrm{B}}( a_m^{\mathrm{in}}|a_k^{\mathrm{fin}})} \right]} = \ln{\left[ \frac{\displaystyle{\widetilde{\sum_{\ell}}\left|m_{km}^{(\ell)}(t)\right|^2}}{e^{\Delta\Phi_{\pi}(k,m)} \displaystyle{\widetilde{\sum_{\ell}}\left|m_{km}^{(\ell)}(t)\right|^2}} \right]} = - \Delta \Phi_{\pi}(k,m)\,,
\end{equation}
where one has defined $\Delta\Phi(k,m) \equiv \Phi_{\pi}(k) - \Phi_{\pi}(m)$ and the summation with the tilde runs only over the Kraus operators (labelled by $\ell$) that include this transition.
Therefore, under the validity of the aforementioned assumptions (i.e., {\rm (i)} $E_{\ell}(t) = \sum_{ij}m_{ji}^{(\ell)}(t)|\pi_j\rangle\!\langle\pi_i|$ with $m_{ji}^{(\ell)}(t)=0\,\,\,\forall t$ if $\Phi_{\pi}(j)-\Phi_{\pi}(i) \neq \Delta \Phi_{\pi}(\ell)$, and {\rm (ii)} $[\mathcal{O}_{\rmi},\pi]=[\mathcal{O}_{\rmf},\pi]=0$), the stochastic quantum entropy production $\Delta\sigma(a_k^{\mathrm{fin}}, a_m^{\mathrm{in}})$ does not depend on the details of the evolution, but it just depends on measurement probabilities evaluated at the initial and final times of the process and on the eigenvalues $\pi_i$, with $i\in \{1,\ldots,d\}$, of the invariant state $\pi$. In particular, one has
\begin{equation}\label{eq:final_stoc_entropy}
    \Delta\sigma(a_k^{\mathrm{fin}}, a_m^{\mathrm{in}}) = - \ln\left[\frac{p(a_k^{\mathrm{fin}})}{p(a_m^{\mathrm{in}})}\right] - \Delta \Phi_{\pi}(k,m) = \Delta s(k,m) - \Delta \Phi_{\pi}(k,m)\,,
\end{equation}
where $\Delta s(k,m) \equiv s(a_k^{\mathrm{fin}}) - s(a_m^{\mathrm{in}})$, with $s(a_m^{\mathrm{in}}) \equiv - \ln p(a_m^{\mathrm{in}})$ and $s(a_k^{\mathrm{fin}}) \equiv - \ln p(a_k^{\mathrm{fin}})$. Thus, $\Delta s(k,m)$ denotes the difference of self-information of measuring the initial and final quantum observables, $\mathcal{O}_{\rm in}$ and $\mathcal{O}_{\rm fin}$ respectively, by recording the corresponding outcomes $a_k^{\mathrm{fin}}$ and $a_m^{\mathrm{in}}$. Let us observe that $\Delta s$ and $\Delta\Phi_{\pi}$ have opposite sign; this indicates that, unlike the unital case, in non-unital quantum maps the total amount of entropy production in a single trajectory can be decreased during the process, e.g., by a mechanism with feedback aka a Maxwell demon with efficacy greater than unity. 

It is worth stressing the role played by the assumptions {\rm (i)} and {\rm (ii)} in achieving Eq.~(\ref{e_SE_noneq}). Assumption {\rm (i)} is responsible for having a single value $\Delta\Phi_{\pi}(\ell)$ for the change of nonequilibrium potential in correspondence of the $\ell^{\rm th}$ Kraus operator $E_{\ell}$; it leads to the validity of the commutation relations (\ref{e_comm_Epi1})-(\ref{eq:commutatore}). Instead, assumption {\rm (ii)} allows to express the ratio $\ln{\left[ \frac{p_{\mathrm{F}}(a_k^{\mathrm{fin}}| a_m^{\mathrm{in}})}{p_{\mathrm{B}}( a_m^{\mathrm{in}}|a_k^{\mathrm{fin}})} \right]}$ in terms of the value of $\Delta\Phi_{\pi}$ that corresponds to the transition $(k,m)$, with $k$ and $m$ labelling the measurement outcomes of $\mathcal{O}_{\rm fin}$ and $\mathcal{O}_{\rm in}$ respectively.

Eq.~(\ref{eq:final_stoc_entropy}) straightforwardly implies the first main result of this paper, which is the fluctuation relation
\begin{equation}\label{eq:DBE}
\frac{
p_{\mathrm{F}}(a_k^{\mathrm{fin}}, a_m^{\mathrm{in}})
}{
p_{\mathrm{B}}(a_m^{\mathrm{in}},a_k^{\mathrm{fin}})
}
= e^{\Delta\sigma(a_k^{\mathrm{fin}},a_m^{\mathrm{in}})} =
e^{\Delta s(k,m)-\Delta \Phi_{\pi}(k,m)}
\end{equation}
describing a symmetry for the ratio of the forward and backward probability distributions of measuring the $(k,m)^{\rm th}$ pair of outcomes from $\mathcal{O}_{\rm in}$ and $\mathcal{O}_{\rm fin}$. 
Specifically, the right-hand-side of Eq.~(\ref{eq:DBE}) depends on the measured outcomes at the initial and final times of the process, and on the invariant state $\pi$ of the non-unital quantum map. Other details of the map are completely absent in Eqs.~(\ref{eq:final_stoc_entropy}) and (\ref{eq:DBE}), albeit we have previously determined that assumptions {\rm (i)}-{\rm (ii)} have to be satisfied in order to get such equations. Moreover, another consequence of Eq.~(\ref{eq:DBE}) is that the identity $p_{\mathrm{F}}(a_k^{\mathrm{fin}},a_m^{\mathrm{in}})=p_{\mathrm{B}}(a_m^{\mathrm{in}},a_k^{\mathrm{fin}})$, which entails vanishing entropy production at the single trajectory level, meaning (strict) reversibility, is valid if and only if 
\begin{equation}\label{eq:reversibility}
\Delta s(k,m)=\Delta \Phi_{\pi}(k,m)\,\,\,\,\forall\,(k,m)\,.
\end{equation}
This is a very strong constraint which holds true when the initial state is the invariant state of the dynamics, so that there is no real evolution (the two measurements have no effect either, because we are assuming that the measured observables commute with the invariant state).

Before concluding this discussion, we comment on a result in Ref.~\cite{ManzanoPRX2018} that looks analogous to Eq.~(\ref{eq:final_stoc_entropy}), and we explain why it is in fact a different result. In Ref.~\cite{ManzanoPRX2018}, the authors assume that they can access and measure both the system (with two observables identified by the projectors $\lbrace\mathcal{P}_n \rbrace, \, \lbrace \mathcal{P}_m^{*} \rbrace$, in the notation of the cited reference) and the reservoir (with two observables having projectors $ \lbrace \mathcal{Q}_{\nu}\rbrace, \, \lbrace \mathcal{Q}_{\mu}^{*} \rbrace $). Each trajectory in their framework is thus specified by both the initial and final outcomes for the system and the initial and final outcomes for the reservoir. There, through the measurements performed on the reservoir, a \textit{unique} Kraus operator, say $M_{\mu \nu}$, is selected for each trajectory of the system conditioned evolution. The latter is indeed described by a quantum operation reading $\mathcal{E}_{\mu \nu}(\cdot) = M_{\mu \nu} (\cdot) M_{\mu \nu}^{\dagger}$.  As a consequence, when computing the forward and backward joint probabilities, a \textit{single} Kraus operator is involved, and thus a single value of the nonequilibrium potential is selected. Hence, without considering any further assumption, one can derive an expression for the entropy production that formally reads as our Eq.~(\ref{eq:final_stoc_entropy}), although each single trajectory carries also the information on the reservoir outcomes. At variance with the setting of Ref.~\cite{ManzanoPRX2018}, we consider the case where only the system can be accessed, and, consistently with such a description, each trajectory is determined by the initial and final outcomes for the system only. We are thus grouping together different reservoir trajectories of Ref.~\cite{ManzanoPRX2018}, so that many Kraus operators are involved in the computation of the joint probabilities Eqs.~(\ref{eq:forward_joint}) and (\ref{eq:backward_joint}). Therefore, in general it is not possible to extract a single value of the nonequilibrium potential for each trajectory, this preventing one to derive Eq.~(\ref{eq:final_stoc_entropy}). The novelty of our result is that, even in the case where one cannot associate a single Kraus operator to a trajectory (e.g., by only knowing the reduced dynamics instead that the conditioned evolution), the fluctuation relation can be restored by choosing a specific set of observables.
It is indeed the measurement that in our case selects a \emph{single transition} for any trajectory and allows the derivation of our result.

\subsection{Average entropy production in terms of quantum relative entropies}

Now, we are going to provide the expression of the average entropy production solely in terms of quantum relative entropies. This is a consequence of the simple expression we found for the stochastic quantum entropy production $\Delta\sigma$ (Eq.~(\ref{eq:final_stoc_entropy})).

To this end, let us consider the probability distribution of $\Delta\sigma$,
\begin{equation}
\mathrm{Prob}(\Delta \sigma) = \sum_{k, m}\delta[\Delta \sigma - \Delta \sigma(a_k^{\mathrm{fin}},a_{m}^{\mathrm{in}}) ]p_{\mathrm{F}}(a_{k}^{\mathrm{fin}}, a_{m}^{\mathrm{in}}),
\end{equation}
where $\delta[\cdot]$ denotes the Dirac delta. Hence, the average of $\Delta\sigma$ reads as $\braket{\Delta \sigma} = \sum_{m,k}\Delta\sigma(a_{k}^{\mathrm{fin}},a_{m}^{\mathrm{in}})p_{\mathrm{F}}(a_k^{\mathrm{fin}},a_m^{\mathrm{in}})$, whereby -- after some calculations -- it can be also written as
\begin{equation}\label{eq:average_entropy}
    \braket{\Delta\sigma} = 
    S(\rho_{\tau}||\rho_{\mathrm{fin}}) + S(\rho_{\mathrm{in}}||\pi) - S(\rho_{\tau}||\pi) \geq 0,
\end{equation}
where $S(\rho||\rho') \equiv \mathrm{Tr}[\rho\ln(\rho) - \rho\ln(\rho')]$ denotes the quantum relative entropy, and the density operators $\rho_{\mathrm{in}}$, $\rho_{\mathrm{fin}}$ and $\rho_{\tau}$ are defined above in Eq.~(\ref{eq:forward_process}). This is the second main result of the present paper. A similar expression, i.e.,  $\braket{\Delta\sigma} = S(\rho_{\mathrm{in}}||\pi) - S(\rho_{\tau}||\pi) $, was found in Ref.~\cite{ManzanoPRX2018}, thus without the explicit contribution of the second measurement that is the first term in Eq.~(\ref{eq:average_entropy}). The full derivation of Eq.~(\ref{eq:average_entropy}) is in Appendix A. Note that, as stated in Eq.~(\ref{eq:average_entropy}), $\braket{\Delta\sigma}$ is always non-negative, because the relative entropy is always non-negative and it is decreasing under the action of CPTP maps on both arguments (therefore the difference of the second and third term is non-negative).

It is worth observing that, if the invariant state of the non-unital quantum map $\Lambda_t$ is the thermal state
\begin{equation}\label{eq:gibbs}
\rho_{\beta} \equiv \frac{e^{-\beta H}}{\mathrm{Tr}[e^{-\beta H}]}
\end{equation}
with $H$ the system Hamiltonian, then the difference of the nonequilibrium potential is $\Delta\Phi_{\pi}(k,m) = \beta(\mathcal{E}_k^{(\pi)}-\mathcal{E}_m^{(\pi)})$, where the $\mathcal{E}_i$'s ($i\in\{1,\ldots,d\}$) denote the possible energy values of the quantum system. i.e. $H\ket{\pi_i}=\mathcal{E}_{i}^{(\pi)}\ket{\pi_i}$. Also, for a thermal invariant state (this is the case we are going to study in Sec.~\ref{sec:qubit_thermalization}), Eq.~(\ref{eq:average_entropy}) becomes
\begin{eqnarray}
    \braket{\Delta\sigma} &=&
    S(\rho_{\tau}||\rho_{\mathrm{fin}})+ S(\rho_{\mathrm{in}}||\rho_{\beta})-S( \rho_{\tau} ||\rho_{\beta}) \nonumber \\
    &=& S(\rho_{\tau}||\rho_{\mathrm{fin}})+ S(\rho_\tau) - S(\rho_{\mathrm{in}}) -  \beta\mathrm{Tr}[H (\rho_\tau - \rho_{\mathrm{in}})],
\end{eqnarray}
where $S(\rho) \equiv -\mathrm{Tr}[\rho \ln(\rho)]$ is the von Neumann entropy. Therefore, one recovers a clear thermodynamic interpretation, with the (average) entropy production corresponding to the (average) entropy variation in the system plus the (average) entropy flux in the thermal bath (that is minus the inverse temperature times the (average) heat exchanged, i.e. $- \beta \Delta Q =: \mathrm{Tr}[H (\rho_\tau - \rho_{\mathrm{in}})]$). In this setting, all the variation of energy is usually attributed to a heat exchange because the Hamiltonian of the system is taken time-independent. It would be interesting to generalize the theory so as to include the case of a non-unital evolution with time-dependent Hamiltonian. In order to do this, one would need to modify the definition of the backward protocol including a dual map that depends on a instantaneous invariant state, i.e. $\pi_t$ such that $\Lambda_t(\pi_t)=\pi_t$. This is matter for future investigation.

\section{Case-study: Qubit thermalization with non-Markovian transient}\label{sec:qubit_thermalization}

In this Section we consider an example to illustrate our previous findings, that is a non-Markovian thermalizing dynamics for a two-level system (qubit). 

The study of non-Markovian quantum dynamics became an active area of research in the last decade~\cite{Rivas_review14,Breuer_review16,deVega_review17,Chruscinski_review22}. In the following, we adopt the nomenclature and definitions proposed in~\cite{Chruscinski2014degree}. In particular, we consider the intermediate propagators $V_{t,s}$, implicitly defined by $\Lambda_t = V_{t,s} \Lambda_s$, and distinguish different degrees of non-Markovianity based on its properties. If $V_{t,s}$ is completely positive (CP) for any pair $t\geq s\geq0$ the dynamics is called CP-divisible or \emph{Markovian}. If instead the propagators are just positive (P) the dynamics is called P-divisible and a dynamics which is P-divisible but not CP-divisible is classified as \emph{weakly non-Markovian}. We will be more interested in dynamics such that the propagator $V_{t,s}$ is not even positive for some pair $t,s$. In this case the dynamics is not P-divisible and it is called \emph{essentially non-Markovian}.

The divisibility properties of the dynamics (and therefore its degree of non-Markovianity) are better studied by looking at the time-dependent generator $\mathcal{L}_t= (\partial_t \Lambda_t ) \Lambda_t^{-1}$, which exists provided the dynamical map $\Lambda_t$ is invertible and its time-derivative is well-defined.

In the following, we consider the prototypical example of a thermalizing dynamics for a qubit, which is described by the time-dependent generator $\mathcal{L}_t$ explicitly written as
\begin{equation}\label{eq:generator}
    \mathcal{L}_t (\cdot) = -i \left[ \frac{\omega}{2} \hat{\sigma}^z,\, \cdot \, \right] +\gamma_\beta(t) \mathrm{e}^{\beta \omega} \left( \hat{\sigma}^- (\cdot) \hat{\sigma}^+ - \frac{1}{2}\Big\{ \hat{\sigma}^+ \hat{\sigma}^-, \, \cdot \, \Big\}   \right) + \gamma_\beta(t) \left( \hat{\sigma}^+ (\cdot) \hat{\sigma}^- - \frac{1}{2}\Big\{ \hat{\sigma}^- \hat{\sigma}^+, \, \cdot \, \Big\}   \right).
\end{equation}
Here $\hat{\sigma}^h$ with $h \in \{ x,y,z \}$ are the usual Pauli matrices, $\hat{\sigma}^\pm = (\hat{\sigma}^x \pm \hat{\sigma}^y)/2$, $\beta$ is the inverse temperature, $\omega$ is the qubit energy gap and $\gamma_\beta(t)$ is a time-dependent decay rate. Using the Bloch vector representation of the qubit density matrix, i.e. $\rho(t)=\frac{1}{2}(\mathbb{I}+\vec{r}(t) \cdot \vec{\hat{\sigma}})$, with $\vec{\hat{\sigma}}\equiv\left\lbrace \hat{\sigma}^x, \hat{\sigma}^y,\hat{\sigma}^z\right\rbrace$ and $\vec{r}(t)\equiv(x(t),y(t),z(t))$, one can rewrite the differential equation $\partial_t \rho(t)= \mathcal{L}_t \rho(t)$ as a system of differential equations for the Bloch vector components
\begin{eqnarray}
&& \partial_t \big(x(t) + i y(t)\big)= \Big(+ i \omega - \frac{(1+\mathrm{e}^{\beta \omega})}{2}\gamma_\beta(t)\Big) \big( x(t) + i y(t)\big), \label{eq:derxy+_t} \\
&& \partial_t \big(x(t) - i y(t)\big)= \Big(-i \omega - \frac{(1+\mathrm{e}^{\beta \omega})}{2}\gamma_\beta(t)\Big) \big( x(t) - i y(t)\big), \label{eq:derxy-_t} \\
&& \partial_t z(t)=- \gamma_\beta(t) (1+\mathrm{e}^{\beta \omega}) z(t) + \gamma_\beta(t) (1-\mathrm{e}^{\beta \omega}) . \label{eq:derz_t}
\end{eqnarray}
The equations above, complemented with the initial conditions $\vec{r}(0)\equiv(x_0,y_0,z_0)$, can be readily solved, so that one can write the Bloch vector components at arbitrary time $t\geq 0$ as follows,
\begin{eqnarray}
&& x(t) + i y(t)= e^{-\Gamma_\beta(t) + i \omega t} \big( x_0 + i y_0 \big), \label{eq:xy+_t} \\
&& x(t) - i y(t)= e^{-\Gamma_\beta(t) - i \omega t} \big( x_0 - i y_0 \big), \label{eq:xy-_t} \\
&& z(t)=e^{-2\Gamma_\beta(t)}z_0 - (e^{-2\Gamma_\beta(t)}-1)z_{\infty} = e^{-2\Gamma_\beta(t)}(z_0-z_{\infty})+z_{\infty}, \label{eq:z_t}
\end{eqnarray}
where we conveniently defined the integrated decay rate $\Gamma_\beta(t) \equiv \frac{1}{2}(1+ \mathrm{e}^{\beta \omega})\int_{0}^{t} \gamma_\beta(\tau) d\tau$ and the parameter $z_{\infty} \equiv - \tanh(\frac{\beta \omega }{2})$. From Eqs.~(\ref{eq:xy+_t})-(\ref{eq:z_t}) one can verify that the unique invariant state for this dynamics is given by the Bloch vector $ \vec{r}_\beta\equiv(0,0,z_\infty) $, representing the Gibbs state $\rho_\beta$ (see Eq.~(\ref{eq:gibbs})) with Hamiltonian $H= \omega/2\, \hat{\sigma}^z$. We assume in the following that the time-dependent rate converges to a positive constant for long times, $\gamma_\beta(t) \overset{t \to \infty}{\longrightarrow} \gamma_\beta >0$, such that the Gibbs state $\rho_\beta$ is also the unique asymptotic state for any initial density matrix. In this sense, the dynamics really describes the thermalization of a qubit to a thermal state, and the parameter $z_\infty$ represents the asymptotic value of $z(t)$, as the notation suggests. 
Note that with the special choice $\gamma_\beta(t)= \lambda(t)  n_\beta = \lambda(t) \frac{1}{\mathrm{e}^{\beta \omega}-1}$ one recovers a time-dependent generalization of the usual quantum optical master equation.
However, we prefer to leave $\gamma_\beta(t)$ unspecified, since the following discussion does not depend qualitatively on this choice. In particular, at the end we may want to compare our findings with the results obtained in Ref.~\cite{Gherardini2020irreversibility} for the unital case (infinite temperature, $\beta=0$), but the quantum optical master equation is not well-defined in this limit, so that another choice of $\gamma_\beta(t)$ will be used.

\subsection{Kraus representation of the quantum map}

In order to show that the dynamics described by Eq.~(\ref{eq:generator}) does satisfy assumption (i), one has to find a Kraus representation corresponding to that evolution. This can be done considering the general expression $\rho(t)\equiv\Lambda_t[\rho(0)]=\sum_{j,k=0}^{3}\lambda_{jk}(t)\hat{\sigma}_j \rho(0)\hat{\sigma}_k$, where $\hat{\sigma}_{\ell}\in\vec{\hat{\sigma}}$ for $\ell \in\{0,\ldots,3\}$, and $\lambda_{ij}(t)$ are 16 complex parameters, substituting the Bloch vector decomposition of $\rho(0)$ and $\rho(t)$, and using the algebraic properties of Pauli matrices to rewrite triples of sigmas into a single sigma. This allows to find a relation between the $\lambda_{ij}(t)$ and the Bloch vector components $x(t),y(t),z(t)$. 
The functions of time $x(t)$, $y(t)$ and $z(t)$ are closely related to the parameters $\lambda_{ij}(t)$ and they can be written as a function of them, as shown in Appendix B.  As provided in Appendix C, the Kraus representation of the quantum map $\Lambda_t$ governing the thermalization of the qubit (with non-Markovian transient) can be achieved by expressing coefficients $\lambda_{ij}(t)$ as a function of the model parameters, i.e., $\Gamma_\beta(t)$, $\omega$ and $\beta$. In this way, one can get the Kraus operators $E_{\ell}$ describing the quantum map in diagonal form $\Lambda_t=\sum_{\ell=1}^{4}E_{\ell}(t)(\cdot)E_{\ell}^{\dagger}(t)$. Specifically, we obtain (see Appendix C for the complete derivation)
\begin{eqnarray}
E_{1} &=& \sqrt{\frac{1}{2}(1-e^{-2\Gamma_\beta(t)})(1+z_{\infty})} \hat{\sigma}^{+} \\
E_{2} &=& \sqrt{\frac{1}{2}(1-e^{-2\Gamma_\beta(t)})(1-z_{\infty})} \hat{\sigma}^{-} \\
E_{3} &=& \sqrt{D_1} \left( u_{1}^{(1)}|0\rangle\!\langle 0|+u_{1}^{(2)}|1\rangle\!\langle 1|\right) \\
E_{4} &=& \sqrt{D_2} \left( u_{2}^{(1)}|0\rangle\!\langle 0|+u_{2}^{(2)}|1\rangle\!\langle 1|\right),
\end{eqnarray}
where $\hat{\sigma}^{+} \equiv |0\rangle\!\langle 1|$, $\hat{\sigma}^{-}\equiv(\hat{\sigma}^{+})^{\dagger}=|1\rangle\!\langle 0|$ and 
\begin{equation*}
D_{1,2}=\frac{1}{2}\left(1+e^{-2\Gamma_\beta(t)}\right)\pm\sqrt{z_{\infty}^2\left(\frac{1-e^{-2\Gamma_\beta(t)}}{2}\right)^2+e^{-2\Gamma_\beta(t)}}\,.
\end{equation*}
Moreover, $u_{i}^{(j)}$ denotes the $j^{\rm th}$ element of the vector $\vec{u}_{i}$, with $i,j=1,2$, where $\vec{u}_{1}\equiv\frac{1}{\sqrt{2b}}\left(-e^{i\omega t}\sqrt{a+b}, \sqrt{b - a}\right)^{T}$ and $\vec{u}_{2}\equiv \frac{1}{\sqrt{2b}} \left(e^{i\omega t}\sqrt{b-a}, \sqrt{a + b}\right)^{T}$, with $a \equiv z_{\infty}(1-e^{-2\Gamma_\beta(t)})/2$ and $b \equiv \sqrt{z_{\infty}^{2}(1-e^{-2\Gamma_\beta(t)})^2+4e^{-2\Gamma_\beta(t)}}/2$.

\subsection{Stochastic entropy production}

In this subsection we evaluate the expressions of the stochastic entropy production and the corresponding statistics for a qubit interacting with a thermal bath in accordance with Eqs.~(\ref{eq:xy+_t})-(\ref{eq:z_t}) entering in $\rho(t)=(\mathbb{I}+\vec{r}(t) \cdot \vec{\hat{\sigma}})/2$. The invariant state of the corresponding map is $\pi=e^{-\beta H}/ \mathrm{Tr}[e^{-\beta H}]$ with $H=\omega \hat{\sigma}^{z}/2$.

Our starting point is the derivation of the stochastic quantum entropy production as defined by Eqs.~\eqref{e_stoc_QEP} and \eqref{e_SE_noneq}. The set of projectors $\left\lbrace \Pi_{m}^{\rmi} \right\rbrace$ and $\left\lbrace \Pi_{k}^{\rmf} \right\rbrace$, associated to the observables $\mathcal{O}_{\rmi}$ and $\mathcal{O}_{\rmf}$ respectively, reads as $\Pi^{\rmi}_{0}=\Pi^{\rmf}_{0}= |0\rangle\!\langle 0|$, with outcomes $a_0^{\rmi}, a_0^{\rmf}$, and $\Pi^{\rmi}_{1}=\Pi^{\rmf}_{1}=|1\rangle\!\langle 1|$, with outcomes $a_1^{\rmi}, a_1^{\rmf}$. Thus, the changes of the nonequilibrium potential, $\Delta \Phi_{\pi}(k,m) = -\ln{\pi_k}+\ln{\pi_m}$, are equal to 
\begin{eqnarray}
&& \Delta\Phi_{\pi}(0,0)=\Delta \Phi_{\pi}(1,1)=0 \\
&& \Delta\Phi_{\pi}(0,1)=-\Delta \Phi_{\pi}(1,0)=\beta\omega \,,
\end{eqnarray}
where we have substituted the expressions of $\pi_0 = e^{-\beta\omega/2}/[2 \cosh{(\beta \omega /2)}]$ and $\pi_1 = e^{\beta \omega / 2}/[2 \cosh{(\beta \omega /2)}]$.

Since for the chosen CPTP maps the equations of motion of $x(t)\pm i y(t)$ decouple from the one of $z(t)$, when fixing $x_{0}=y_{0}=0$ the dynamical evolution (after the $1^{\rm st}$ measurement of the TPM scheme) remains diagonal with respect to the $\hat{\sigma}^{z}$ basis. Hence, the state $\rho_{\rmi}$ is evolved at time $\tau$ as $\rho_{\tau}=\Lambda_{\tau}(\Pi_m^{\rmi})= \frac{1}{2}[\mathbb{I}+z_m(\tau)\hat{\sigma}^{z}]$, with $z_{m}(\tau) \equiv e^{-2\Gamma_\beta(\tau)}[z_m(0)-z_{\infty}]+z_{\infty}$ where $z_{m}(0) = \pm 1$ for $m = 0,1$ (we recall that $m$ is the label of the initial measurement outcomes). Let us note that the outcome of the second projective measurement is $a_0^{\rmf}$, with probability $p(a_0^{\rmf})=[1+z_m(\tau)]/2$, and $a_1^{\rmf}$, with $p(a_1^{\rmf})=[1-z_m(\tau)]/2$. 
As a result, the values of the stochastic entropy production $\Delta\sigma(k,m)$, labelled by the index over the measurement outcomes, are provided by the following quantities:
\begin{equation}
\begin{split}
& \Delta \sigma (0,0) = \ln \left[ \frac{p(a_0^{\rmi})}{[1+z_0(\tau)]/2} \right], \quad \Delta \sigma (0,1) = \ln \left[ \frac{1-p(a_0^{\rmi})}{[1+z_1(\tau)]/2} \right] -\beta \omega,\\ 
& \Delta \sigma (1,0) = \ln \left[ \frac{p(a_0^{\rmi})}{[1-z_0(\tau)]/2} \right] + \beta \omega, \quad \Delta \sigma (1,1) = \ln \left[ \frac{1-p(a_0^{\rmi})}{[1-z_1(\tau)]/2} \right].
\end{split}
\end{equation}
In order to derive the probability distribution of the stochastic entropy production, as well as the corresponding statistical moments, we need to evaluate for the forward process the conditional probability that the outcome $a_k^{\rmf}$ occurs after the outcome $a_m^{\rmi}$ from the first measurement, i.e., $p_{\mathrm{F}}(a_k^{\rmf}|a_{m}^{\rmi})=\tr[\Pi_k^{\rmf} \Lambda_{\tau} (\Pi_{m}^{\rmi})]$. Considering the four possible combinations of outcomes, the conditional probabilities are
\begin{equation}
\begin{split}
& p_{\mathrm{F}}(a_0^{\rmf}|a_{0}^{\rmi})=\tr[\Pi_0^{\rmf} \Lambda_{\tau} (\Pi_{0}^{\rmi})]= \frac{1}{2}[1+z_0(\tau)], \qquad p_{\mathrm{F}}(a_0^{\rmf}|a_{1}^{\rmi})=\tr[\Pi_0^{\rmf} \Lambda_{\tau} (\Pi_{1}^{\rmi})]= \frac{1}{2}[1+z_1(\tau)], \\
& p_{\mathrm{F}}(a_1^{\rmf}|a_{0}^{\rmi})=\tr[\Pi_1^{\rmf} \Lambda_{\tau} (\Pi_{0}^{\rmi})]= \frac{1}{2}[1-z_0(\tau)], \qquad p_{\mathrm{F}}(a_1^{\rmf}|a_{1}^{\rmi})=\tr[\Pi_1^{\rmf} \Lambda_{\tau} (\Pi_{1}^{\rmi})]= \frac{1}{2}[1-z_1(\tau)].
\end{split}
\end{equation}

Now we can provide the expression of the average entropy production. In doing this, for the sake of an easier presentation, we assume that the initial state of the forward process at time $t=0$ is one of the eigenstates of $\hat{\sigma}^{z}$, i.e., $\rho(0)=\frac{1}{2}[\mathbb{I} + z_m(0) \hat{\sigma}^{z}]$.
Hence, the state after the first measurement prescribed by the TPM scheme remains unchanged: $\rho_{\rmi} = \rho(0)= \frac{1}{2}[\mathbb{I} + z_m(0) \hat{\sigma}^{z}] = \Pi_m^{\rmi}$ such that $p(a_0^{\rmi})=1$. As a result, one gets:
\begin{equation}\label{e_mv_p0in_1}
\braket{\Delta\sigma} = -\left\lbrace \frac{1+z_0(\tau)}{2}\ln  \left[ \frac{1+z_0(\tau)}{2} \right] + \frac{1-z_0(\tau)}{2} \ln  \left[ \frac{1-z_0(\tau)}{2} \right] \right\rbrace + \beta\omega \left[ \frac{1-z_0(\tau)}{2} \right],
\end{equation}
which allows us to derive a relatively simple expression for its time-derivative:
\begin{equation}\label{e_dermv}
\partial_t \braket{\Delta \sigma} = - \frac{1}{2}\partial_t z_0 (t) \left\lbrace \ln \left[ \frac{1+z_0(t)}{1-z_0(t)}\right] + \beta\omega \right\rbrace.
\end{equation}
The full derivation of $\braket{\Delta \sigma}$ by initializing the system in a generic density operator $\rho_0$ is in Appendix D. 

Always under the assumption of $p(a_0^{\rmi})=1$, we can also derive the second statistical moment of the stochastic entropy production, i.e.,   
\begin{eqnarray}
\braket{\Delta \sigma ^2} & = & \frac{1+z_0(\tau)}{2}\ln^2  \left[ \frac{1+z_0(\tau)}{2} \right] + \frac{1-z_0(\tau)}{2} \ln^2  \left[ \frac{1-z_0(\tau)}{2} \right]\nonumber \\
&+& \beta^2\omega^2 \frac{(1-z_0(\tau))}{2} -2\beta\omega \frac{(1-z_0(\tau))}{2}\ln \left[ \frac{1-z_0(\tau)}{2}\right],
\end{eqnarray}
and the corresponding variance:
\begin{equation}\label{eq:variance_Delta_sigma}
\mathrm{Var}(\Delta \sigma)=\braket{\Delta \sigma ^2}-\braket{\Delta \sigma}^2 = \frac{1}{4}[1-z_0^2(t)]\left\lbrace \ln \left[ \frac{1+z_0(t)}{1-z_0(t)}\right] +\beta\omega\right\rbrace^{2}.
\end{equation}
In Appendix D the reader can also find the analytical expression of the time-derivative of $\mathrm{Var}(\Delta\sigma)$ that will be employed in the next Section about irreversibility mitigation.  

\section{Irreversibility mitigation}
\label{sec:irreversibility_mitigation}

As already discussed in the unital case, we want to find a connection between the non-Markovianity of the dynamics and the phenomenon we called \emph{irreversibility mitigation}~\cite{Gherardini2020irreversibility}. More explicitly, we look for time intervals in which both the average entropy production and the variance are decreasing. In the perfectly reversible case, which is zero entropy production in a single trajectory, the distribution $\mathrm{Prob}(\Delta \sigma)$ is a Dirac delta in zero. Since the dynamics we consider are irreversible, the average of the distribution typically shifts towards positive values as time passes and the variance broadens (one no-longer has a delta distribution). Therefore, irreversibility mitigation stems for the fact that, due to non-Markovianity, in a certain time-interval the distribution $\mathrm{Prob}(\Delta \sigma)$ tends to get closer to a delta in zero.

We consider the dynamics studied in the previous Section and we choose the initial state to be $|0\rangle\!\langle 0|$ as before, so that the dynamics is completely described by the $z$ component of the Bloch vector. 
From Eqs.~(\ref{eq:xy+_t})-(\ref{eq:z_t}), using $z_0=1$ one can compute
\begin{equation}
 \partial_t z(t)= -2 \,\partial_t \Gamma_\beta(t) \, \mathrm{e}^{-2 \Gamma_\beta(t)} (1-z_{\infty}),
\end{equation}
so that the sign of $\partial_t z(t)$ is opposite to the sign of $\partial_t \Gamma_\beta(t)= \frac{1}{2}(1+e^{\beta \omega}) \gamma_\beta(t)$. It is known that the sign of $\gamma_\beta(t)$ is related to the divisibility property of the dynamics and therefore to its degree of non-Markovianity (see the discussion at the beginning of the previous Section). In particular, the dynamics is CP-divisible and P-divisible (in this particular example the two notions coincide) if and only if $\gamma_\beta(t) \geq 0$ for any time $t$. This can be seen for instance from Example 3 in~\cite{Chruscinski2014degree}.

In order to determine the sign of $\partial_t \braket{\Delta \sigma} $, one has to study the sign of the quantity in brackets in \eqref{e_dermv}, that we call $I(t)$ in the following. It is possible to show that it is always non-negative in a few steps
\begin{align}\label{e_It}
   I(t)= &\ln \left[ \frac{1+z(t)}{1-z(t)}\right] + \beta \omega = \nonumber \\
    =&\ln \left[ \frac{1+z(t)}{1-z(t)}\right] - \ln \left[ \frac{1+z_\infty}{1-z_\infty}\right] = \nonumber\\
    =&\ln \left[ \frac{1+z(t)}{1+z_\infty}\right] - \ln \left[ \frac{1-z(t)}{1-z_\infty}\right] = \nonumber\\
    =&\ln \left[ \frac{1+z_\infty + \mathrm{e}^{-2 \Gamma_\beta(t)}(1-z_\infty) }{1+z_\infty}\right] - \ln \left[ \frac{1-z_\infty - \mathrm{e}^{-2 \Gamma_\beta(t)}(1-z_\infty)}{1-z_\infty}\right] = \nonumber\\
    =&\ln \left( 1 + \mathrm{e}^{-2 \Gamma_\beta(t)} \mathrm{e}^{\beta \omega}  \right) - \ln \left( 1 - \mathrm{e}^{-2 \Gamma_\beta(t)}  \right) \nonumber \\
    =&\ln \left( \frac{1 + \mathrm{e}^{-2 \Gamma_\beta(t)} \mathrm{e}^{\beta \omega} }{1 - \mathrm{e}^{-2 \Gamma_\beta(t)}}\right) \geq 0 .
\end{align}
Therefore, for $\beta$ positive, the sign of $\partial_t \braket{\Delta \sigma} $ corresponds to the sign of $\gamma_\beta(t)$ and, as a consequence, the average entropy production is decreasing whenever the dynamics is not P-divisible (essentially non-Markovian). This is consistent with the general result obtained in~\cite{Marcantoni2017entropy}, valid for arbitrary initial state.

Let us now focus on the derivative of the variance. We rewrite the last line of equation \eqref{e_dervar} for convenience
\begin{equation}
 \partial_t \mathrm{Var}(\Delta \sigma) = 2 \partial_t \braket{\Delta \sigma}\left( \frac{z(t)}{2} I(t) -1 \right).
\end{equation}
The interest is in time intervals such that both the variance and the average are decreasing. Therefore, in particular, their derivatives have to share the same sign. This means the term in parenthesis has to be positive.
A necessary but not sufficient condition is $z(t)\geq 0$. Indeed, if this is not true the term is evidently negative. More explicitly, the condition reads
\begin{equation}
  \Gamma_\beta(t) \leq -\frac{1}{2}\ln\left(\frac{1- \mathrm{e}^{-\beta \omega}}{2}\right).
\end{equation}

By means of a somewhat lengthy calculation that we present in Appendix E, one can also find a sufficient condition to produce irreversibility mitigation:
\begin{equation}
    \Gamma_\beta(t) \leq -\frac{1}{2}\ln(x_+) , \quad\quad x_+ = \frac{2}{5}\left(1- \mathrm{e}^{-\beta\omega} + \sqrt{\mathrm{e}^{-2\beta\omega} + 3 \mathrm{e}^{-\beta\omega} +1 }\right)
\end{equation}

In order to make the comparison with the unital case we can consider a $\gamma_\beta(t)$ that admits a finite limit for $\beta=0$. In particular, one can take $\gamma_\beta(t) \equiv \gamma(t)$, independent from $\beta$. In this case one has for infinite temperature $$\Gamma_0(t)\leq -\frac{1}{2}\log\left(\sqrt{\frac{4}{5}}\right) \sim 0.056.$$ In Ref.~\cite{Gherardini2020irreversibility} we found $\Gamma_0(t) \leq 0.091$ for unital maps (note the different notation: $\Gamma_0(t)$ in this paper corresponds to $\Phi(t)$ in Ref.~\cite{Gherardini2020irreversibility}). One could still refine this bound using the theory of Pad\'e approximants~\cite{book_Pade} (see also Appendix E), however, our aim was to show that a parameter range in which irreversibility mitigation happens does exist.

\section{Conclusions}
\label{sec:conclusions}

In this paper we studied the stochastic quantum entropy production in non-unital quantum dynamics (e.g., thermalization).
We started recalling the Tasaki-Crooks formalism to introduce the stochastic entropy production, and then, as a first result, we studied the conditions whereby the latter obeys a fluctuation relation based on the so-called nonequilibrium potential, as in Refs.~\cite{ManzanoPRX2018,Manzano15}. The nonequilibrium potential brings information on the invariant state of the open dynamics under scrutiny, and, as we have shown, it is responsible for the unbalance of the corresponding forward and backward processes. In the case of a thermal invariant state, it has a clear interpretation in terms of energy fluxes. Then, the conditions such that the stochastic entropy production obeys a fluctuation theorem can be summarised as follows: i) the Kraus operators induce jumps with determined energy gaps, and ii) the measured observables commute with the invariant state of the map.
In fact, with respect to Refs.~\cite{ManzanoPRX2018,Manzano15} and in line with Refs.~\cite{GherardiniQST2018,Gherardini2020irreversibility}, we also clarified the role of the chosen measurement scheme (TPM) both in the definition of the stochastic entropy production and in the derivation of its statistics. This led us to write the average entropy production in terms of quantum relative entropies, with a contribution that explicitly depends on the second measurement. This is the second main result.

Subsequently, we recalled the concept of quantum non-Markovianity, by using a thermalizing qubit dynamics with non-Markovian transient as a case-study. Then, as a third novel result, we determined a parameter range (depending on the temperature of the invariant state of the map) allowing for the average entropy production and its variance to be decreasing at the same time in a transient, a phenomenon that we called irreversibility mitigation in~\cite{Gherardini2020irreversibility}.

Let us now discuss some possible outlook of our findings. First of all, our results are based on a construction of the backward protocol starting from the invariant state of the forward dynamics, which is assumed to be unique and strictly positive. It would be interesting to extend the framework in order to include cases with multiple and/or non-invertible invariant states. For instance, in the case of (infinitely) many invariant states one could construct a backward protocol for each of them and then ask if any of those protocols satisfies the condition on the Kraus operators that we used to prove the fluctuation theorem. Also, it would be worth understanding what is the relation between different choices of the reference invariant state. Even more strikingly, when the invariant state is not invertible one has to resort to a different method for building the reversed protocol, like for instance the one proposed in~\cite{Aurell15}. The existence of fluctuation relations in this framework needs to be investigated further. Moreover, even in the setting of this paper, as already argued above, our derivations hold true for non-unital quantum dynamics satisfying certain constraints on the Kraus operators. For the moment, they have been tested on a single case-study provided by qubit thermalization maps with non-Markovian transient. This allowed us to observe how irreversibility mitigation modifies as a function of the temperature of an external bath. However, it would be interesting to study in detail different examples. In particular, it might be worth investigating whether our results apply to quantum maps (with a microscopic derivation) that are customarily considered in the scenario of open quantum systems (i.e. whether they admit the specific Kraus form we considered). Among them, we mention quantum collision models~\cite{CiccarelloReview2022} and quantum Maxwell's demons~\cite{Lin13,Campisi17,NajeraSantosPRR2020,PoulsenPRE2022,HernandezGomezPRXQuantum22} as both deal with repeated interactions with external agents (or just laser pulses) over time, making non-unital the dynamics of the quantum system. Furthermore, one might consider how the nonequilibrium potential changes by employing measurement schemes beyond TPM~\cite{Micadei20,Sone2020quantum,Levy20,Gherardini20x,hernandez2022experimental}, so as to understand the role played by quantum coherence/correlation terms in the initial state. Finally, as a long-term project, one might study if the irreversibility mitigation is favoured or hindered by increasing the size of the open quantum system under scrutiny. In this respect, spin-boson models like the multimode Dicke model could be a viable platform of investigation.

\section*{Acknowlegments}

We acknowledge financial support from The Blanceflor Foundation through the project ``The theRmodynamics behInd thE meaSuremenT postulate of quantum mEchanics (TRIESTE)'', from the MISTI Global Seed Funds MIT-FVG Collaboration Grant ``Non-Equilibrium Thermodynamics of Dissipative Quantum Systems (NETDQS)'', from the EPSRC through the Grant No. EP/R04421X/1, from the European Research Council through the ERC Starting Grant MaMBoQ (No. 802901), from the PNRR MUR project PE0000023-NQSTI and from the QUARESC project PID2019-109094GB-C21 (AEI /10.13039/501100011033). This work was performed under the auspices of GNFM-INDAM.

\section*{Data availability statement} Our manuscript has no associated data.

\section*{Author contribution statement}

S.M. and S.G. conceived the main ideas of the project. E.F. performed most of the calculations with inputs from S.M. and S.G. All the authors discussed the results and contributed to writing and revising the manuscript.

\section*{Conflict of interest statement}

The authors declare that they have no conflicts of interest.

\section*{APPENDICES}

\setcounter{figure}{0}
\setcounter{table}{0}

\makeatletter
\renewcommand{\bibnumfmt}[1]{[#1]}
\renewcommand{\citenumfont}[1]{#1}

\subsection*{Appendix A: Average entropy production in terms of quantum relative entropies: Formal derivation}

Let us compute the average value of the stochastic quantum entropy production,  which reads as
\begin{eqnarray}\label{e_mv_sigma}
\braket{\Delta \sigma} &=&  \sum_{m,k} \Delta \sigma(a_{k}^{\mathrm{fin}},a_{m}^{\mathrm{in}})p_{\rm F}(a_k^{\mathrm{fin}}|a_m^{\mathrm{in}})p(a_m^{\mathrm{in}})\nonumber \\
&=& \sum_{m,k} \left\lbrace \ln{[p(a_m^{\mathrm{in}})]}- \ln{[p(a_k^{\mathrm{fin}})] } -\Delta \Phi_{\pi}(k,m) \right\rbrace \mathrm{Tr}[\Pi_k^{\mathrm{fin}}\Lambda_{\tau}(\Pi_m^{\mathrm{in}})]p(a_m^{\mathrm{in}}) \,.
\end{eqnarray}
In Eq.~\eqref{e_mv_sigma} we can consider separately two contributions: we start by evaluating
\begin{eqnarray}
&&\displaystyle{\sum_{m,k} \left\lbrace \ln{[p(a_m^{\mathrm{in}})]}- \ln{[p(a_k^{\mathrm{fin}})] }\right\rbrace \mathrm{Tr}[\Pi_k^{\mathrm{fin}}\Lambda_{\tau}(\Pi_m^{\mathrm{in}})]p(a_m^{\mathrm{in}}) }\nonumber \\
&=&\displaystyle{ \mathrm{Tr}\left\lbrace \sum_k \Pi_k^{\mathrm{fin}} \Lambda_{\tau}\left[ \sum_m \Pi_m^{\mathrm{in}} p(a_m^{\mathrm{in}})\ln{[p(a_m^{\mathrm{in}})]} \right] \right\rbrace - \mathrm{Tr}\left\lbrace \sum_k \Pi_k^{\mathrm{fin}} \ln{[p(a_k^{\mathrm{fin}})]}\Lambda_{\tau}\left[ \sum_m \Pi_m^{\mathrm{in}} p(a_m^{\mathrm{in}}) \right] \right\rbrace } \nonumber \\
&=& \displaystyle{ \mathrm{Tr}\left[ 
\Lambda_{\tau}(\rho_{\mathrm{in}} \ln{\rho_{\mathrm{in}}}) \right] - \mathrm{Tr}\left[ \ln{\rho_{\mathrm{fin}}} \Lambda_{\tau}(\rho_{\mathrm{in}}) \right]= \mathrm{Tr}\left[ \rho_{\mathrm{in}} \ln{\rho_{\mathrm{in}}} \right] - \mathrm{Tr}\left[ \rho_{\tau} \ln{\rho_{\mathrm{fin}}}\right] +\mathrm{Tr}\left[ \rho_{\tau} \ln{\rho_{\tau}} \right] - \mathrm{Tr}\left[ \rho_{\tau} \ln{\rho_{\tau}} \right] } \nonumber \\
&=& \displaystyle{ \Delta S_{(\rho_{\tau}, \rho_{\mathrm{in}})} + S(\rho_{\tau}|| \rho_{\mathrm{fin}}) }\,, 
\end{eqnarray}
where $ \Delta S_{(\rho, \rho')} \equiv S(\rho)-S(\rho')$, with $S(\rho) \equiv -\mathrm{Tr}[\rho \ln(\rho)]$ the Von Neumann entropy, and $S( \rho || \rho') \equiv \mathrm{Tr}[\rho \ln(\rho)- \rho \ln(\rho')]$ is the quantum relative entropy. Instead, the second contribution to Eq.~\eqref{e_mv_sigma} reads
\begin{eqnarray}
&& -\sum_{m,k} \Delta\Phi_{\pi}(k,m) \mathrm{Tr}[\Pi_k^{\mathrm{fin}}\Lambda_{\tau}(\Pi_m^{\mathrm{in}})]p(a_m^{\mathrm{in}})\nonumber \\ 
&& =\mathrm{Tr}\left\lbrace\sum_{k}\Pi_k^{\mathrm{fin}}\ln[\pi_k]\,\Lambda_{\tau}\left[\sum_{m}\Pi_m^{\mathrm{in}}p(a_m^{\mathrm{in}})\right]\right\rbrace 
- \mathrm{Tr}\left\lbrace\sum_{k}\Pi_k^{\mathrm{fin}}\Lambda_{\tau}\left[\sum_{m}\Pi_m^{\mathrm{in}}p(a_m^{\mathrm{in}})\ln[\pi_m]\right]\right\rbrace\nonumber \\
&&=\mathrm{Tr}\left[\rho_{\tau}\ln{\pi}\right] - \mathrm{Tr}\left[\rho_{\mathrm{in}}\ln\pi\right]\,.
\end{eqnarray}
Putting together the two contributions, one finally gets
\begin{eqnarray}
\braket{\Delta \sigma} &=& S(\rho_{\tau}|| \rho_{\mathrm{fin}}) + S(\rho_{\tau}) - S(\rho_{\mathrm{in}})+ 
\mathrm{Tr}[\rho_{\tau}\ln{\pi}]
-\mathrm{Tr}[\rho_{\mathrm{in}}\ln{\pi}] \nonumber \\
&=& S(\rho_{\tau}||\rho_{\mathrm{fin}}) 
+ S(\rho_{\mathrm{in}}||\pi)-S(\rho_{\tau}||\pi)\,,
\end{eqnarray}
as provided by Eq.~(\ref{eq:average_entropy}) in the main text. 

\subsection*{Appendix B: Generic open dynamics of a qubit in Kraus representation}

As shown in the main text, the time evolution of the density operator of a qubit subjected to a generic open CPTP map $\Lambda_{t}$ can be written both as $\rho(t)=\frac{1}{2}(\mathbb{I}+\vec{r}(t) \cdot \vec{\hat{\sigma}})$ and as $\rho(t)=\sum_{j,k=0}^{3}\lambda_{jk}(t)\hat{\sigma}_j \rho(0)\hat{\sigma}_k$, where $\vec{r}(t)$, $\vec{\hat{\sigma}}$, $\lambda_{jk}(t)$ and $\hat{\sigma}_{\ell}$, with $j,k=0,\ldots,3$ and $\ell=0,\ldots,3$, are defined in the main text.
The parameters $\lambda_{ij}(t)$ obey the following constraints:
\begin{itemize}
\item[•] Unit-trace preservation: 
\begin{itemize}
\item[(a)] $ \mathrm{Re[\lambda_{01}(t)]}= \mathrm{Im[\lambda_{32}(t)]}$, $ \mathrm{Re[\lambda_{02}(t)]}= \mathrm{Im[\lambda_{13}(t)]}$, $ \mathrm{Re[\lambda_{03}(t)]}= \mathrm{Im[\lambda_{21}(t)]}$
\item[(b)] $\sum_{i=0}^{3} \lambda_{ii}(t)=1$
\end{itemize} 
\item[•] Hermiticity preservation: $\lambda_{jk}^{*}(t)=\lambda_{kj}(t)$ for any $k,j=0,\ldots,3$ such that
\begin{itemize}
\item[(c)] $\lambda_{ii}(t)$ are real functions for any $t$ and $i=0,\ldots,3$
\item[(d)] $\mathrm{Re}[\lambda_{ij}]=\mathrm{Re}[\lambda_{ji}]$ and $\mathrm{Im}[\lambda_{ij}]=-\mathrm{Im}[\lambda_{ji}]$ for $i \neq j$
\end{itemize}  
\item[•] ${\bf \lambda} \equiv [\lambda_{ij}(t)]_{ij}$ positive matrix $\iff$ the quantum map $\Lambda_t$ is CP.
\end{itemize}
Moreover, the relation between the vector $\vec{r}(t)$ and the parameters $\lambda_{ij}(t)$ 
reads as follows:
\begin{equation}\label{e_xvsl}
\begin{split} 
x(t) = & 
4 \mathrm{Re}[\lambda_{01}(t)]+x_{0}[\lambda_{00}(t)+\lambda_{11}(t)-\lambda_{22}(t)-\lambda_{33}(t)]+y_{0}\lbrace -2\mathrm{Im}[\lambda_{03}(t)]+2\mathrm{Re}[\lambda_{21}(t)]\rbrace + \\ 
& z_0\lbrace 2\mathrm{Im}[\lambda_{02}(t)] + 2\mathrm{Re}[\lambda_{31}(t)]\rbrace \, ,
\end{split}
\end{equation}
\begin{equation}\label{e_yvsl}
\begin{split}  
y(t) = & 4 \mathrm{Re}[\lambda_{02}(t)]+y_{0}[1-2\lambda_{11}(t)-2\lambda_{33}(t)]+x_{0}\lbrace 2\mathrm{Im}[\lambda_{03}(t)]+2\mathrm{Re}[\lambda_{12}(t)]\rbrace + \\ & 
z_0\lbrace -2\mathrm{Im}[\lambda_{01}(t)] + 2\mathrm{Re}[\lambda_{13}(t)]\rbrace \, ,
\end{split}
\end{equation}
\begin{equation}\label{e_zvsl}
\begin{split} 
z(t) = & 
4 \mathrm{Re}[\lambda_{03}(t)]+z_0[1-2\lambda_{11}(t)-2\lambda_{22}(t)]+x_{0}\lbrace -2\mathrm{Im}[\lambda_{02}(t)]+2\mathrm{Re}[\lambda_{13}(t)]\rbrace + \\ & 
y_{0}\lbrace 2\mathrm{Im}[\lambda_{01}(t)] + 2\mathrm{Re}[\lambda_{32}(t)]\rbrace\,.
\end{split}
\end{equation}
Then, as discussed in the main text, the coefficients $\lambda_{ij}(t)$, which overall are 16 complex coefficients, have to obey specific constraints that allows for the quantum map to be physical, namely to always return a proper density operator for any value of time $t$.

\subsection*{Appendix C: Kraus representation of the map governing qubit thermalization with non-Markovian transient}

The first step consists in deriving the coefficients $\lambda_{ij}(t)$ as a function of the parameters of the model, and thus in obtaining the (non-diagonal) Kraus representation of the quantum map $\Lambda_t$. 
From Eq.\,\eqref{e_zvsl} and Eq.\,\eqref{eq:z_t}, one gets
\begin{equation}
\begin{split}
e^{-2\Gamma_\beta(t)}z_0-(e^{-2\Gamma_\beta(t)}-1)z_{\infty} = & 4 \mathrm{Re}[\lambda_{03}(t)]+z_0[1-2\lambda_{11}(t)-2\lambda_{22}(t)]+\\
& x_0\lbrace -2\mathrm{Im}[\lambda_{02}(t)]+2\mathrm{Re}[\lambda_{13}(t)]\rbrace + y_0\lbrace 2\mathrm{Im}[\lambda_{01}(t)] + 2\mathrm{Re}[\lambda_{32}(t)]\rbrace .
\end{split}
\end{equation}
This equality has to be satisfied for arbitrary $x_0,y_0$ and $z_0$. As a consequence one obtains four constraints, which are derived equating the coefficients of the independent terms.
Explicitly one has
\begin{itemize}
\item[(I)] 
$ \mathrm{Re}[\lambda_{03}(t)] = \frac{z_{\infty}}{4} (1-e^{-2\Gamma_\beta(t)})$, with $\mathrm{Re}[\lambda_{03}(t)]\underbrace{=}_{(d)} \mathrm{Re}[\lambda_{30}(t)] \underbrace{=}_{(a)}  \mathrm{Im}[\lambda_{21}(t)] \underbrace{=}_{(d)} - \mathrm{Im}[\lambda_{12}(t)]$
\item[(II)] $1-2\lambda_{11}(t)-2\lambda_{22}(t) = e^{-2\Gamma_\beta(t)}$
\item[(III)] $-2\mathrm{Im}[\lambda_{02}(t)]+2\mathrm{Re}[\lambda_{13}(t)]=0$, so that $-\mathrm{Im}[\lambda_{20}(t)] \underbrace{=}_{(d)} \mathrm{Im}[\lambda_{02}(t)] \underbrace{=}_{\mathrm{(III)}} \mathrm{Re}[\lambda_{13}(t)] \underbrace{=}_{(d)} \mathrm{Re}[\lambda_{31}(t)]$
\item[(IV)] 
$2\mathrm{Im}[\lambda_{01}(t)] + 2\mathrm{Re}[\lambda_{32}(t)]=0$, entailing also $-\mathrm{Im}[\lambda_{10}(t)] \underbrace{=}_{(d)} \mathrm{Im}[\lambda_{01}(t)] \underbrace{=}_{\mathrm{(IV)}} -\mathrm{Re}[\lambda_{32}(t)] \underbrace{=}_{(d)} -\mathrm{Re}[\lambda_{23}(t)]$\,,
\end{itemize}
where (a)-(d) are the conditions defined in Appendix B concerning the constraints of unit-trace preservation, Hermiticity preservation and positivity of the matrix ${\bf \lambda}$. Analogously, comparing  Eqs.~\eqref{e_xvsl} and \eqref{e_yvsl} with \eqref{eq:xy-_t} and \eqref{eq:xy+_t}, we get 
\begin{equation}
\begin{split}
x(t) \pm i y(t) = &  e^{-\Gamma_\beta(t) \pm i \omega t} [ x_{0} \pm i y_{0}] = e^{-\Gamma_\beta(t)}\cos(\omega t) [x_{0}\pm i y_{0}] + e^{-\Gamma_\beta(t)}\sin(\omega t)[\pm i x_{0} - y_{0}] = \\
&  4 \mathrm{Re}[\lambda_{01}(t)]+x_{0}[\lambda_{00}(t)+\lambda_{11}(t)-\lambda_{22}(t)-\lambda_{33}(t)]-y_{0}\lbrace 2\mathrm{Im}[\lambda_{03}(t)]-2\mathrm{Re}[\lambda_{21}(t)]\rbrace + \\
& z_0\lbrace 2\mathrm{Im}[\lambda_{02}(t)] + 2\mathrm{Re}[\lambda_{31}(t)]\rbrace \pm i 4 \mathrm{Re}[\lambda_{02}(t)]\pm iy_{0}[1-2\lambda_{11}(t)-2\lambda_{33}(t)] \pm \\
& i x_{0}\lbrace 2\mathrm{Im}[\lambda_{03}(t)]+2\mathrm{Re}[\lambda_{12}(t)]\rbrace  \pm i z_0\lbrace -2\mathrm{Im}[\lambda_{01}(t)] + 2\mathrm{Re}[\lambda_{13}(t)]\rbrace,
\end{split}
\end{equation}
which implies the following constraints when comparing term by term
\begin{itemize}
\item[(V)] $\mathrm{Re}[\lambda_{01}(t)] = 0$, 
entailing that $\mathrm{Re}[\lambda_{10}(t)] \underbrace{=}_{(d)} \mathrm{Re}[\lambda_{01}(t)] \underbrace{=}_{(a)}  \mathrm{Im}[\lambda_{32}(t)]\underbrace{=}_{(d)} - \mathrm{Im}[\lambda_{23}(t)] =0$ 
%
\item[(VI)] $\mathrm{Re}[\lambda_{02}(t)] = 0$,
such that $\mathrm{Re}[\lambda_{20}(t)] \underbrace{=}_{(d)} \mathrm{Re}[\lambda_{02}(t)] \underbrace{=}_{(a)}  \mathrm{Im}[\lambda_{13}(t)]\underbrace{=}_{(d)} - \mathrm{Im}[\lambda_{31}(t)] =0$ 
%
\item[(VII)] $2\mathrm{Im}[\lambda_{02}(t)] + 2\mathrm{Re}[\lambda_{31}(t)]=0$, 
from which $\mathrm{Im}[\lambda_{02}(t)]=-\mathrm{Re}[\lambda_{31}(t)]\underbrace{=}_{\mathrm{(III)}} - \mathrm{Im}[\lambda_{02}(t)]$, implying
$\mathrm{Im}[\lambda_{02}(t)]=\mathrm{Im}[\lambda_{20}(t)]=\mathrm{Re}[\lambda_{13}(t)] = \mathrm{Re}[\lambda_{31}(t)]= 0$
%
\item[(VIII)] $-2\mathrm{Im}[\lambda_{01}(t)] + 2\mathrm{Re}[\lambda_{13}(t)] = 0$,
meaning that $\mathrm{Im}[\lambda_{01}(t)]=\mathrm{Re}[\lambda_{13}(t)] \underbrace{=}_{\mathrm{(VII)}} 0$. Thus, similarly to the previous case, one has
$\mathrm{Im}[\lambda_{01}(t)]=\mathrm{Im}[\lambda_{10}(t)] =\mathrm{Re}[\lambda_{23}(t)]=\mathrm{Re}[\lambda_{32}(t)] = 0$
%
\item[(IX)] 
$2\mathrm{Im}[\lambda_{03}(t)]+2\mathrm{Re}[\lambda_{12}(t)]=2\mathrm{Im}[\lambda_{03}(t)]-2\mathrm{Re}[\lambda_{21}(t)]= e^{-\Gamma_\beta(t)}\sin(\omega t)$,  
so that
\begin{itemize}
\item[IX.a] $-\mathrm{Im}[\lambda_{30}(t)] \underbrace{=}_{(d)} \mathrm{Im}[\lambda_{03}(t)] = \frac{1}{2} e^{-\Gamma_\beta(t)}\sin(\omega t)$
\item[IX.b] 
$\mathrm{Re}[\lambda_{12}(t)] = - \mathrm{Re}[\lambda_{21}(t)] \underbrace{=}_{(d)} -\mathrm{Re}[\lambda_{12}(t)]$, 
Hence, $\mathrm{Re}[\lambda_{21}(t)] = \mathrm{Re}[\lambda_{12} (t)] = 0$
\end{itemize}
\item[(X)] 
$\lambda_{00}(t)+\lambda_{11}(t)-\lambda_{22}(t)-\lambda_{33}(t) = 1-2\lambda_{11}(t)-2\lambda_{33}(t) = e^{-\Gamma_\beta(t)}\cos(\omega t)$ that entails
\begin{itemize}
\item[X.a]  $\lambda_{11}=\lambda_{22}= \frac{1}{4}(1-e^{-2\Gamma_\beta(t)})$ that is obtained by also exploiting (b) and (II)
\item[X.b] 
$\lambda_{33}(t) = \frac{1}{4}(1+e^{-2\Gamma_\beta(t)})+\frac{1}{2}e^{-\Gamma_\beta(t)}\cos(\omega t)$ 
\item[X.c]
$\lambda_{00}(t) = \frac{1}{4}(1+e^{-2\Gamma_\beta(t)})-\frac{1}{2}e^{-\Gamma_\beta(t)}\cos(\omega t)$\,.
\end{itemize}
\end{itemize}
In conclusion, the coefficients $\lambda_{i,j}(t)$ read as
\begin{itemize}
    \item 
    $\lambda_{11}=\lambda_{22}= \frac{1}{4}(1-e^{-2\Gamma_\beta(t)}) \equiv \lambda(t)$
    so that $e^{-\Gamma_\beta(t)}=\sqrt{1-4\lambda(t)}$
    \item 
    $\lambda_{00}(t) = \frac{1}{4}(1+e^{-2\Gamma_\beta(t)})-\frac{1}{2}e^{-\Gamma_\beta(t)}\cos(\omega t) = \frac{1}{2}-\lambda(t) -\frac{1}{2}e^{-\Gamma_\beta(t)}\cos(\omega t)$
    \item 
    $\lambda_{33}(t) = \frac{1}{4}(1+e^{-2\Gamma_\beta(t)})+\frac{1}{2}e^{-\Gamma_\beta(t)}\cos(\omega t) = \frac{1}{2}-\lambda(t) +\frac{1}{2}e^{-\Gamma_\beta(t)}\cos(\omega t)$
    \item 
    $\mathrm{Re}[\lambda_{30}(t)]= \mathrm{Re}[\lambda_{03}(t)] =  \mathrm{Im}[\lambda_{21}(t)] = - \mathrm{Im}[\lambda_{12}(t)] = \frac{z_{\infty}}{4} (1-e^{-2\Gamma_\beta(t)}) = z_{\infty} \lambda(t)$
    \item 
    $-\mathrm{Im}[\lambda_{30}(t)] \underbrace{=}_{(d)} \mathrm{Im}[\lambda_{03}(t)] = \frac{1}{2} e^{-\Gamma_\beta(t)}\sin(\omega t)$,
\end{itemize}
and all the other coefficients vanish. Accordingly, the Kraus representation of the map $\Lambda_{t}$ is 
\begin{equation}\label{equation_map_app}
\begin{split}
\Lambda_t(\cdot) = & 
+ i\mathrm{Re}[\lambda_{30}(t)]\left[\hat{\sigma}_{2}(\cdot)\hat{\sigma}_{1}-\hat{\sigma}_{1}(\cdot)\hat{\sigma}_{2}\right] 
+ \lambda_{11}(t)\left[\hat{\sigma}_{1}(\cdot)\hat{\sigma}_{1}+\hat{\sigma}_{2}(\cdot)\hat{\sigma}_{2}\right]
+\mathrm{Re}[\lambda_{03}(t)]\left[ \hat{\sigma}_{0}(\cdot)\hat{\sigma}_{3}+\hat{\sigma}_{3}(\cdot)\hat{\sigma}_{0}\right] + \\
& i\mathrm{Im}[\lambda_{03}(t)]\left[\hat{\sigma}_{0}(\cdot)\hat{\sigma}_{3}-\hat{\sigma}_{3}(\cdot)\hat{\sigma}_{0}\right]+\lambda_{00}(t)\hat{\sigma}_{0}(\cdot)\hat{\sigma}_{0}+\lambda_{33}(t)\hat{\sigma}_{3}(\cdot)\hat{\sigma}_{3}=\\
& = + i z_{\infty}\lambda(t)\left[\hat{\sigma}_{2}(\cdot)\hat{\sigma}_{1}-\hat{\sigma}_{1}(\cdot)\hat{\sigma}_{2}\right]+ \lambda(t) \left[\hat{\sigma}_{1}(\cdot)\hat{\sigma}_{1}+\hat{\sigma}_{2}(\cdot)\hat{\sigma}_{2}\right]+z_{\infty}\lambda(t)\left[ \hat{\sigma}_{0}(\cdot)\hat{\sigma}_{3}+\hat{\sigma}_{3}(\cdot)\hat{\sigma}_{0}\right] + \\
& i\frac{1}{2}e^{-\Gamma_\beta(t)}\sin(\omega t)\left[\hat{\sigma}_{0}(\cdot)\hat{\sigma}_{3}-\hat{\sigma}_{3}(\cdot)\hat{\sigma}_{0}\right] +\left[\frac{1}{2}-\lambda(t)\right] \left[ \hat{\sigma}_{0}(\cdot)\hat{\sigma}_{0} +  \hat{\sigma}_{3}(\cdot)\hat{\sigma}_{3}\right] +\frac{1}{2}e^{-\Gamma_\beta(t)}\cos(\omega t) \left[ \hat{\sigma}_{3}(\cdot)\hat{\sigma}_{3}-\hat{\sigma}_{0}(\cdot)\hat{\sigma}_{0}\right]=\\
&=  2\lambda(t)(1+z_{\infty})\hat{\sigma}^{+}(\cdot)\hat{\sigma}^{-}+2\lambda(t)(1-z_{\infty})\hat{\sigma}^{-}(\cdot)\hat{\sigma}^{+}+2\left[\frac{1}{2}-\lambda(t)+z_{\infty}\lambda(t)\right]|0\rangle\!\langle 0|(\cdot)|0\rangle\!\langle 0| + \\ & 2\left[\frac{1}{2}-\lambda(t)-z_{\infty}\lambda(t)\right]|1\rangle\!\langle 1|(\cdot)|1\rangle\!\langle 1|-e^{-\Gamma_\beta(t)}e^{i\omega t }|0\rangle\!\langle 0|(\cdot)|1\rangle\!\langle 1|-e^{-\Gamma_\beta(t)}e^{-i\omega t }|1\rangle\!\langle 1|(\cdot)|0\rangle\!\langle 0|
\end{split}
\end{equation}
with $\hat{\sigma}^{+} \equiv |0\rangle\!\langle 1|$ and $\hat{\sigma}^{-}\equiv(\hat{\sigma}^{+})^{\dagger}=|1\rangle\!\langle 0|$. 
We can now write the quantum map $\Lambda_t$ in the diagonal form $\Lambda_t(\cdot)= \sum_{\ell} E_{\ell}(t) (\cdot) E_{\ell}^{\dagger}(t)$. In fact, from the first line in the last step of Eq.~\eqref{equation_map_app}, we can already identify two diagonal operators, i.e.,
\begin{eqnarray}
& E_{1} \equiv \sqrt{2\lambda(t)(1+z_{\infty})} \hat{\sigma}^{+} = \sqrt{\frac{1}{2}(1-e^{-2\Gamma_\beta(t)})(1+z_{\infty})} \hat{\sigma}^{+} \\
& E_{2} \equiv \sqrt{2\lambda(t)(1-z_{\infty})} \hat{\sigma}^{-} = \sqrt{\frac{1}{2}(1-e^{-2\Gamma_\beta(t)})(1-z_{\infty})} \hat{\sigma}^{-}\,,
\end{eqnarray}
where one can verify that $2\lambda(t)(1 \pm z_{\infty}) \geq 0$, as $\Gamma_\beta(t) \geq 0$ and $-1 \leq \pm z_{\infty} \leq 1$. In order to get also the operators $E_{3}$ and $E_{4}$ we diagonalize the remaining part of the map $\Lambda_t$ that reads as $\sum_{i,j=1}^{2}c_{ij}V_{i}(\cdot)V_{j}^{\dagger}$, where 
$V_{k}\in\{|0\rangle\!\langle 0|,|1\rangle\!\langle 1|\}$ and $c_{ij}$ are the elements of the matrix 
\begin{equation}
C=\begin{pmatrix}
2\left[\frac{1}{2}-\lambda(t)+z_{\infty}\lambda(t)\right] && -e^{-\Gamma_\beta(t)}e^{i\omega t} \\
-e^{-\Gamma_\beta(t)}e^{-i\omega t} && 2\left[\frac{1}{2}-\lambda(t)-z_{\infty}\lambda(t)\right]
\end{pmatrix}.
\end{equation}
Thus, if we denote $D$ and $U$ the diagonalized matrix containing the eigenvalues of $C$ and $U$ the unitary operator that diagonalizes (i.e., spectrally decomposes) $C$ so that $C = UDU^{\dagger}$, then the two remaining Kraus operators are respectively equal to $E_{3}=\sqrt{D_{1}}\sum_{i=1,2}U_{i1}V_{i}$ and $E_{4}=\sqrt{D_{2}}\sum_{i=1,2}U_{i2}V_{i}$. The eigenvalues $D_{1}$ and $D_{2}$ of the matrix $C$ are
\begin{equation}
D_{1,2}= 2\left[\frac{1}{2}-\lambda(t)\right]\pm2\sqrt{[z_{\infty}\lambda(t)]^2+\left[\frac{1}{4}-\lambda(t)\right]} = \frac{1}{2}\left(1+e^{-2\Gamma_\beta(t)}\right)\pm\sqrt{z_{\infty}^2\left(\frac{1-e^{-2\Gamma_\beta(t)}}{2}\right)^2+e^{-2\Gamma_\beta(t)}}\,,
\end{equation}
and the corresponding eigenvectors read as
\begin{eqnarray}
&& \vec{u}_{1} = \frac{1}{\sqrt{2b}}\left(-e^{i\omega t}\sqrt{a+b}, \sqrt{b-a}\right)^{T}\\
&& \vec{u}_{2}= \frac{1}{\sqrt{2b}} \left(e^{i\omega t}\sqrt{b-a}, \sqrt{a+b}\right)^{T},
\end{eqnarray}
where we have defined $a \equiv z_{\infty}(1-e^{-2\Gamma_\beta(t)})/2$ and $b \equiv \sqrt{z_{\infty}^{2}(1-e^{-2\Gamma_\beta(t)})^2+4e^{-2\Gamma_\beta(t)}}/2$, with the properties $(a + b)(\pm a \mp b) = \mp e^{-2\Gamma_\beta(t)} $
As a result, the additional Kraus operators have the following expressions:
\begin{eqnarray}
& E_{3} = \sqrt{D_1} \left( u_{1}^{(1)}|0\rangle\!\langle 0|+u_{1}^{(2)}|1\rangle\!\langle 1|\right) \\
& E_{4} = \sqrt{D_2} \left( u_{2}^{(1)}|0\rangle\!\langle 0|+u_{2}^{(2)}|1\rangle\!\langle 1|\right), 
\end{eqnarray}
where $u_{i}^{(j)}$ denotes the $j^{\rm th}$ element of the vector $\vec{u}_{i}$ with $i,j=1,2$.

\subsection*{Appendix D: Qubit thermalization: Average entropy production \& $2^{\rm nd}$ statistical moment of $\Delta\sigma$}

In this Appendix we report the complete derivation of both the average entropy production and the second statistical moment of $\Delta\sigma$
for the considered case-study of the qubit thermalization with non-Markovian transient. These calculations are carried out by considering a generic initial state $\rho_0$ but still under the assumption to apply the TPM scheme when defining the probability distribution of the stochastic entropy production. The notation and all the symbols appearing in the equations below have been already introduced both in the main text of the paper and in the previous Appendices.   

Explicitly, the mean value of the stochastic entropy production is provided by
\begin{eqnarray}
\braket{\Delta\sigma} & = &  \sum_{k,m} \Delta \sigma (a_k^{\rmf}, a_m^{\rmi})\tr[\Pi_k^{\rmf} \Lambda_{\tau} (\Pi_{m}^{\rmi})] p(a_m^{\rmi}) \nonumber \\
& = & \ln \left\lbrace \frac{p(a_0^{\rmi})}{[1+z_0(\tau)]/2} \right\rbrace \frac{1}{2}[1+z_0(\tau)] p(a_0^{\rmi})+ \left\lbrace \ln \left[ \frac{1-p(a_0^{\rmi})}{[1+z_1(\tau)]/2} \right] -\beta \omega \right\rbrace \frac{1}{2}[1+z_1(\tau)][1-p(a_0^{\rmi})]\nonumber \\
& + & \left\lbrace \ln \left[ \frac{p(a_0^{\rmi})}{[1-z_0(\tau)]/2} \right] + \beta \omega  \right\rbrace\frac{1}{2}[1-z_0(\tau)]p(a_0^{\rmi})+\ln \left\lbrace \frac{1-p(a_0^{\rmi})}{[1-z_1(\tau)]/2} \right\rbrace\frac{1}{2}[1-z_1(\tau)][1-p(a_0^{\rmi})] \nonumber \\
& = & p(a_0^{\rmi}) \ln [p(a_0^{\rmi})] + [1-p(a_0^{\rmi})] \ln [1-p(a_0^{\rmi})] \nonumber \\
& - & p(a_0^{\rmi})\left\lbrace \frac{1+z_0(\tau)}{2}\ln  \left[ \frac{1+z_0(\tau)}{2} \right] + \frac{1-z_0(\tau)}{2} \ln  \left[ \frac{1-z_0(\tau)}{2} \right] \right\rbrace \nonumber \\
& - & [1-p(a_0^{\rmi})]  \left\lbrace \frac{1+z_1(\tau)}{2} \ln  \left[ \frac{1+z_1(\tau)}{2} \right]+ \frac{1-z_1(\tau)}{2} \ln  \left[ \frac{1-z_1(\tau)}{2} \right]\right\rbrace \nonumber  \\
& - &\beta \left\lbrace \frac{\omega}{2} \left[ p(a_0^{\rmi})z_0(\tau)+ (1- p(a_0^{\rmi}))z_1(\tau)  \right] - \frac{\omega}{2}\left[ 2p(a_0^{\rmi})-1 \right] \right\rbrace .
\end{eqnarray}
Furthermore, the $2^{\rm nd}$ statistical moment of $\Delta\sigma$ reads as
\begin{eqnarray}
\braket{\Delta\sigma^2} & = & 
\ln^2 \left\lbrace \frac{p(a_0^{\rmi})}{[1+z_0(\tau)]/2} \right\rbrace \frac{1}{2}[1+z_0(\tau)] p(a_0^{\rmi})+ \left\lbrace \ln \left[ \frac{1-p(a_0^{\rmi})}{[1+z_1(\tau)]/2} \right] -\beta \omega \right\rbrace^2 \frac{1}{2}[1+z_1(\tau)][1-p(a_0^{\rmi})] \nonumber \\
& + & \left\lbrace \ln \left[ \frac{p(a_0^{\rmi})}{[1-z_0(\tau)]/2} \right] + \beta \omega  \right\rbrace^2 \frac{1}{2}[1-z_0(\tau)]p(a_0^{\rmi})+\ln^2 \left\lbrace \frac{1-p(a_0^{\rmi})}{[1-z_1(\tau)]/2} \right\rbrace\frac{1}{2}[1-z_1(\tau)][1-p(a_0^{\rmi})] \nonumber\\
& = & p(a_0^{\rmi}) \ln^2 [p(a_0^{\rmi})] + [1-p(a_0^{\rmi})] \ln^2 [1-p(a_0^{\rmi})] \nonumber \\
& + & p(a_0^{\rmi})\left\lbrace \frac{1+z_0(\tau)}{2}\ln^2  \left[ \frac{1+z_0(\tau)}{2} \right] + \frac{1-z_0(\tau)}{2} \ln^2  \left[ \frac{1-z_0(\tau)}{2} \right] \right\rbrace \nonumber \\
& + & [1-p(a_0^{\rmi})]  \left\lbrace \frac{1+z_1(\tau)}{2} \ln^2  \left[ \frac{1+z_1(\tau)}{2} \right]+ \frac{1-z_1(\tau)}{2} \ln^2  \left[ \frac{1-z_1(\tau)}{2} \right]\right\rbrace \nonumber  \\
& - & 2p(a_0^{\rmi})\ln[p(a_0^{\rmi})]\left\lbrace \frac{1+z_0(\tau)}{2}\ln  \left[ \frac{1+z_0(\tau)}{2} \right] + \frac{1-z_0(\tau)}{2} \ln  \left[ \frac{1-z_0(\tau)}{2} \right] \right\rbrace \nonumber \\
& - & 2[1-p(a_0^{\rmi})]\ln[1-p(a_0^{\rmi})]  \left\lbrace \frac{1+z_1(\tau)}{2} \ln  \left[ \frac{1+z_1(\tau)}{2} \right]+ \frac{1-z_1(\tau)}{2} \ln  \left[ \frac{1-z_1(\tau)}{2} \right]\right\rbrace \nonumber \\
& + &\beta^2\omega^2 \left\lbrace p(a_{0}^{\rmi}) \frac{1-z_0(\tau)}{2}+ [1-p(a_{0}^{\rmi})]\frac{1+z_1(\tau)}{2} \right\rbrace \nonumber \\
& + & 2\beta \omega \left\lbrace  p(a_0^{\rmi})\ln[p(a_0^{\rmi})] \frac{1-z_0(\tau)}{2} - [1-p(a_0^{\rmi})]\ln[1-p(a_0^{\rmi})] \frac{1+z_1(\tau)}{2} \right. \nonumber \\
& + & \left. [1-p(a_0^{\rmi})]\frac{1+z_1(\tau)}{2}\ln\left[ \frac{1+z_1(\tau)}{2}\right] - p(a_0^{\rmi})\frac{1-z_0(\tau)}{2}\ln \left[ \frac{1-z_0(\tau)}{2} \right] \right\rbrace. 
\end{eqnarray}
Given Eq.~(\ref{eq:variance_Delta_sigma}) for the entropy variance in the main text, valid for $p(a_0^{\rmi})=1$, we also write explicitly here the time-derivative of $\mathrm{Var}(\Delta\sigma)$: 
\begin{eqnarray}\label{e_dervar}
\partial_t \mathrm{Var}(\Delta\sigma) &=&  -\frac{1}{2}z_0(t)\partial_t z_0(t) \left\lbrace \ln \left[ \frac{1+z_0(t)}{1-z_0(t)}\right] +\beta \omega	 \right\rbrace^2 \nonumber \\
&+& \frac{1}{4} \cancel{[1-z^2_0(t)]}2\left\lbrace \ln \left[ \frac{1+z_0(t)}{1-z_0(t)}\right] +\beta \omega	 \right\rbrace \cancel{\left[ \frac{1-z_0(t)}{1+z_0(t)} \right]} \frac{[1-\bcancel{z_0(t)}]\partial_t z_{0}(t)-[1+\bcancel{z_0(t)}](-\partial_t z_{0}(t))}{\cancel{[1-z_{0}(t)]^2}}\nonumber \\
&=& -\frac{1}{2}z_0(t)\partial_t z_0(t) \left\lbrace \ln \left[ \frac{1+z_0(t)}{1-z_0(t)}\right] +\beta \omega	 \right\rbrace^2 + \left\lbrace \ln \left[ \frac{1+z_0(t)}{1-z_0(t)}\right] +\beta \omega	 \right\rbrace \partial_t z_0(t) \nonumber \\
&=& -\frac{1}{2}\partial_t z_0(t) \left\lbrace \ln \left[ \frac{1+z_0(t)}{1-z_0(t)}\right] +\beta \omega	 \right\rbrace \left\lbrace z_0(t)\left[ \ln \left( \frac{1+z_0(t)}{1-z_0(t)}\right) +\beta \omega	\right] -2 \right\rbrace \nonumber \\
&=& \partial_t \braket{\Delta \sigma}\left\lbrace z_0(t)\left[ \ln \left( \frac{1+z_0(t)}{1-z_0(t)}\right) +\beta \omega	\right] -2 \right\rbrace \nonumber \\
&=& 2 \partial_t \braket{\Delta \sigma}\left\lbrace \frac{z_0(t)}{2}\left[ \ln \left( \frac{1+z_0(t)}{1-z_0(t)}\right) +\beta \omega	\right] -1 \right\rbrace.
\end{eqnarray}

\subsection*{Appendix E: Bound on the variance}

In order to find a sufficient condition such that $z(t)I(t) \geq 2$, we use the explicit expression for $I(t)$ given in the second to last line of \eqref{e_It} and bound it through the following inequalities~\cite{Topsoe_2007}
\begin{align}
    \frac{2x}{2+x} &\leq \ln(1+x) \leq \frac{x}{2}\left(\frac{2+x}{1+x}\right),  \quad 0 \leq x < \infty \label{eq:log_bound1} \\
    \frac{x}{2}\left(\frac{2+x}{1+x}\right) &\leq  \ln(1+x) \leq \frac{2x}{2+x},  \quad\quad\quad -1 < x \leq 0 . \label{eq:log_bound2}
\end{align}
Explicitly, one has
\begin{equation}
    A_1 \leq \ln \left( 1 + \mathrm{e}^{-2 \Gamma_\beta(t)} \mathrm{e}^{\beta \omega}  \right) \leq B_1, \quad A_2 \leq \ln \left( 1 - \mathrm{e}^{-2 \Gamma_\beta(t)}  \right) \leq B_2, \quad -B_2 \leq - \ln \left( 1 - \mathrm{e}^{-2 \Gamma_\beta(t)}  \right) \leq - A_2
\end{equation}
and as a consequence also
\begin{equation}\label{e_bounds}
    A_1 -B_2 \leq \ln \left( 1 + \mathrm{e}^{-2 \Gamma_\beta(t)} \mathrm{e}^{\beta \omega}  \right)  - \ln \left( 1 - \mathrm{e}^{-2 \Gamma_\beta(t)}  \right)  \leq B_1 - A_2
\end{equation}
with
\begin{align*}
    & A_1 = \frac{2 \,\mathrm{e}^{-2 \Gamma_\beta(t)} \mathrm{e}^{\beta \omega} }{2+ \mathrm{e}^{-2 \Gamma_\beta(t)} \mathrm{e}^{\beta \omega}}, \quad B_1 = \left(\frac{\mathrm{e}^{-2 \Gamma_\beta(t)} \mathrm{e}^{\beta \omega}}{2} \right) \frac{2+ \mathrm{e}^{-2 \Gamma_\beta(t)} \mathrm{e}^{\beta \omega}}{1+ \mathrm{e}^{-2 \Gamma_\beta(t)} \mathrm{e}^{\beta \omega}}, \\ 
    & A_2= \frac{-\mathrm{e}^{-2 \Gamma_\beta(t)}}{2}\left(\frac{2-\mathrm{e}^{-2 \Gamma_\beta(t)}}{1-\mathrm{e}^{-2 \Gamma_\beta(t)}} \right), \quad B_2 = \frac{-2\mathrm{e}^{-2 \Gamma_\beta(t)}}{2-\mathrm{e}^{-2 \Gamma_\beta(t)}}.
\end{align*}
Coming back to $z(t)I(t)$, assuming $z(t)\geq 0$, and using the lower bound in \eqref{e_bounds} one has
\begin{align}
    z(t)I(t) &= \left(\mathrm{e}^{-2 \Gamma_\beta(t)}\Big(1 + \tanh(\beta\omega/2) \Big) - \tanh(\beta\omega/2) \right) \ln \left( \frac{1 + \mathrm{e}^{-2 \Gamma_\beta(t)} \mathrm{e}^{\beta \omega} }{1 - \mathrm{e}^{-2 \Gamma_\beta(t)}}\right) \geq \nonumber \\
    &\geq \left(\mathrm{e}^{-2 \Gamma_\beta(t)}\Big(1 + \tanh(\beta\omega/2) \Big) - \tanh(\beta\omega/2) \right) \frac{4 \mathrm{e}^{-2 \Gamma_\beta(t)} (\mathrm{e}^{\beta\omega} +1) }{\left( 2+ \mathrm{e}^{-2 \Gamma_\beta(t)} \mathrm{e}^{\beta\omega}\right)\left(2- \mathrm{e}^{-2 \Gamma_\beta(t)} \right) } = \nonumber \\
    &= 2 + \frac{10 \mathrm{e}^{\beta\omega} \mathrm{e}^{-4 \Gamma_\beta(t)} - 8(\mathrm{e}^{\beta\omega} -1) \mathrm{e}^{-2 \Gamma_\beta(t)} -8  }{\left( 2+ \mathrm{e}^{-2 \Gamma_\beta(t)} \mathrm{e}^{\beta\omega}\right)\left(2- \mathrm{e}^{-2 \Gamma_\beta(t)} \right)},
\end{align}
so that in order to have $z(t)I(t)-2\geq 0$ it is sufficient to require that the following inequality holds true
\begin{equation}
    10 \mathrm{e}^{\beta\omega} \mathrm{e}^{-4 \Gamma_\beta(t)} - 8(\mathrm{e}^{\beta\omega} -1) \mathrm{e}^{-2 \Gamma_\beta(t)} -8 \geq 0.
\end{equation}
This is a second order algebraic inequality for $x=\mathrm{e}^{-2 \Gamma_\beta(t)}$, which is satisfied when the variable $x$ obeys $x\geq x_+$ or $x\leq x_-$ with $x_\pm$ solutions of the corresponding algebraic equality
\begin{equation}
    x_\pm = \frac{2}{5}\left(1- \mathrm{e}^{-\beta\omega} \pm \sqrt{\mathrm{e}^{-2\beta\omega} + 3 \mathrm{e}^{-\beta\omega} +1 }\right).
\end{equation}
Since $x_-$ is negative the inequality $\mathrm{e^{-2 \Gamma_\beta(t)}} \leq x_-$ is never satisfied and the only sensible constraint is $\mathrm{e^{-2 \Gamma_\beta(t)}} \geq x_+$. The value $x_+$ is always positive and one can show that it is always smaller than $1$ for any positive value of $\beta$. More precisely, it is always between $\sqrt{4/5}$ (value for $\beta=0$) and $4/5$ (value for $\beta \to \infty$) and monotonically decreasing in $\beta$. We have therefore a second bound that is sufficient to produce derivative of the variance and derivative of the average with the same sign
\begin{equation}
    \Gamma_\beta(t) \leq -\frac{1}{2}\ln(x_+) .
\end{equation}
As a consistency check, we can now compare the two bounds and see that the sufficient bound is always stricter than the necessary bound. This can be readily done
\begin{align}
    -\frac{1}{2}\ln\left(\frac{1- \mathrm{e}^{-\beta \omega}}{2}\right) &\geq -\frac{1}{2}\ln(x_+) \nonumber\\
    \frac{1- \mathrm{e}^{-\beta \omega}}{2} &\leq \frac{2}{5}\left(1- \mathrm{e}^{-\beta\omega} + \sqrt{\mathrm{e}^{-2\beta\omega} + 3 \mathrm{e}^{-\beta\omega} +1 }\right) \nonumber\\
    \frac{1- \mathrm{e}^{-\beta \omega}}{10} &\leq \frac{2}{5}\sqrt{\mathrm{e}^{-2\beta\omega} + 3 \mathrm{e}^{-\beta\omega} +1 } \nonumber\\
    \mathrm{e}^{-2\beta\omega} -2\mathrm{e}^{-\beta \omega} +1 &\leq 16 \Big( \mathrm{e}^{-2\beta\omega} + 3 \mathrm{e}^{-\beta\omega} +1 \Big) \nonumber\\
    0 &\leq 15\mathrm{e}^{-2\beta\omega} + 50 \mathrm{e}^{-\beta\omega} +15 \,. 
\end{align}
One could also think of improving the sufficient bound replacing the rational functions in Eqs.~(\ref{eq:log_bound1}),(\ref{eq:log_bound2}) with rational functions of higher order (Pad\'e approximations~\cite{book_Pade}). However, the refined bounds would probably not be particularly enlightening. Our aim was to show that it is indeed possible to have a parameter regime such that irreversibility mitigation happens, and this is what we demonstrated.

\bibliography{Biblio}

\end{document}